\documentclass[%
 reprint,
nofootinbib,
 amsmath,amssymb,
 aps,
pra,
]{revtex4-2}

\usepackage{graphicx}
\usepackage{dcolumn}
\usepackage{bm}

\begin{document}

\author{
V.~V.~Strelkov$^{1,2,
\ast}$
}
\affiliation{
\mbox{$^{1}$Prokhorov General Physics Institute of the Russian Academy of Sciences, Vavilova street 38, Moscow 119991, Russia} \\
\mbox{$^{2}$Institute of Applied Physics of the Russian Academy of Sciences, 46 Ulyanov st., Nizhny Novgorod 603950, Russia} \\
$^{\ast}$strelkov.v@gmail.com
}

\title{%
Dark and bright autoionizing states in resonant high harmonic generation: simulation via 1D helium model
}

\begin{abstract}
 We study the role of dark and bright autoionizing states (AIS) in photoionization and high harmonic generation (HHG) using a 1D helium model. This model allows numerical integration of the time-dependent Schr\"odinger equation beyond the singe-electron approximation completely taking into account electronic correlation. We find the level structure of the system and the spatial distribution of the electronic density for several states including AIS. Studying the HHG efficiency as a function of the detuning from the resonances with AIS we find the HHG enhancement lines. The shapes of these lines are different from the corresponding Fano lines in the photoelectronic spectra, in agreement with the experimental studies in helium.
Moreover, we simulate HHG under the conditions when the fundamental frequency is close to the even-order multiphoton resonance with the dark AIS. We find the enhanced generation of the neighbouring odd harmonics. The details of the enhancement lines for these harmonics can be understood taking into account the temporal delay between the emission of the non-resonant and resonant XUV; this delay is defined by the AIS lifetime. Finally, our simulations show that the HHG enhancement due to the dark and the bright AIS is comparable in the studied system.
\end{abstract}

\maketitle

\section*{Introduction}
Two-electron atoms are probably the simplest systems where the electronic correlations play crucial role. The Schr\"odinger equation describing correlated three-body Coulomb dynamics cannot be resolved analytically; several approximate yet very accurate analytical methods have been developed to describe the helium atom in absence of the external field~\cite{Bete_Salpeter, Review_two-electron_atoms}.  In the presence of the intense laser field the numerical integration of the Schr\"odinger equation is the most effective tool to study the correlated dynamics in a two-electron system. This can be done directly~\cite{Smyth1998} for the 3D helium atom; however, such calculation is feasible for the limited range of the field parameters and anyway it is very computer-demanding (this makes it relevant to consider the simplified 3D two-electron system~\cite{Ruiz2006}). Moreover, many aspects of the multi-electron dynamics can be simulated via 1D helium model~\cite{1D_He_1976, 1d_He_1980, 1D_He_1996, 1D_He_2018}. In particular, it can be used to study such dynamics in the laser field~\cite{Grobe1992, Haan1994, Bauer1997, Lappas_1998, Lein2000, Borisova2020, Volkova2004, Roso2005,Zhao2012, Koval2007, Efimov2018}.

The existence of auto-ionizing states (AIS) (quasi-stable atomic states whose excitation energies exceed the ionization threshold) is one of the striking multi-electronic phenomena. The role of these states in high-order harmonic generation (HHG) is actively studied both theoretically~\cite{Milosevic_2007, Kheifets2008, Strelkov2010, Frolov2010, Strelkov2014, Wahyutama2019} and experimentally~\cite{Ganeev2006, Chang2008, Singh2021, Shiner2011, Fareed2018} (for a review see~\cite{Ganeev2012_Review} and references therein).When the one-photon transition from AIS to the ground state is allowed (such AIS is called a bright one) and the transition frequency is close to the harmonic frequency, the HHG can be strongly enhanced (by more than an order of magnitude). This can be attributed to the AIS population from the continuum in the re-scattering process, and XUV emission via the AI-ground state transition~\cite{Strelkov2010}. Moreover, the HHG enhancement by a bright AIS {\it dressed} by two laser photons was observed~\cite{Fareed2017}.  

If the transition is forbidden the AIS is called a dark one. In particular, this is the case when the AI state has the same parity as the ground one. This AIS can be populated by an {\it even} number of laser photons, but the XUV at the AI-ground state frequency cannot be emitted. However, the dark AIS dressed by the laser photon can absorb~\cite{Magunov2001, Argenti2015} or emit the XUV photon. Very recently the HHG enhancement due to dressed dark AIS was observed in indium plasma plume~\cite{Fareed2022}.

In this paper we study the resonant properties of HHG using the numerical integration of the time-dependent Schr\"odinger equation (TDSE) for 1D helium. We investigate the level structure of the 1D helium atom and ion with an emphasis on the AIS properties. Knowledge of the level structure allows attributing HHG enhancement found in numerical calculations to resonances with certain AI states. In particular, we investigate the enhancement of two neighbouring odd harmonics due to a single resonance with the dark AIS. Calculating intensities and phases of these harmonics we show that the shape of the resonant harmonic enhancement line can be linked to the attosecond properties of the resonant XUV emission. 

\section{Level structure of 1D helium}

We use the following model potential (atomic units are used throughout the paper):
\begin{equation}
V(x,y)=\frac{-2}{\sqrt{x^2+a^2}}+\frac{-2}{\sqrt{y^2+a^2}}+\frac{1}{\sqrt{(x-y)^2+b^2}},
\label{eq_Potential}
\end{equation}
where $x$ and $y$ are the electrons' coordinates, $a$ and $b$ are constants. We use $a=1/\sqrt 2=0.707$ and $b=1/\sqrt 3=0.577$ to reproduce the first and the second ionization potential of the actual helium, see Appendix for more details. Note that the similar value of the latter constant was found in~\cite{Zhao2012}, and the former constant was found in Ref~\cite{Javanainen}. The 2-D time-dependent Schr\"odinger equation (TDSE) for the atomic potential given by Eq.~(\ref{eq_Potential}) is solved numerically with the approach described in \cite{Strelkov_2006}. The solution is done on a spatial grid of $N \times N$ nodes; we use $N$ from 500 to 1000 depending on the conditions; the spatial step is $0.2$; 100 notes at the boundaries of the numerical box are occupied by the (almost non-reflecting) absorbing layers. 

The level structures of the ion and the atom are illustrated by Fig.~\ref{Levels}, the method for calculating the energies of the levels is described in the Appendix. We present both the binding energies of the levels (multiplied by $-1$) and the excitation energies with respect to the atomic ground state. 

\begin{figure}[h]
\centering
\includegraphics[width=0.9\linewidth]{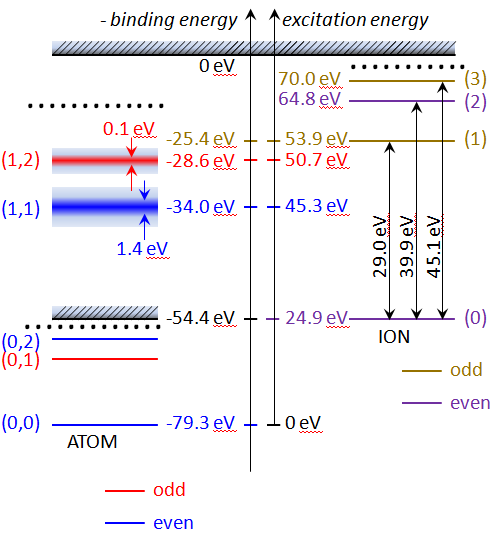}
\caption{The level structure of the model 1D helium atom and the 1D hydrogen-like ion. Zero of the excitation energy is the ground state of the atom; zero of the binding energy is the second ionization threshold.}
\label{Levels}
\end{figure}

An atomic level is characterized by a pair of integer quantum numbers $n_x$ and $n_y$; $n_{x,y}=0,1...$. The state is even if $n_x+n_y$ is even, and odd if $n_x+n_y$ is odd. A quantum number can be understood as describing the state of one electron\footnote{In contrast to the 3D atom, the ground state in one-electronic 1D system is usually denoted by $n=0$ (so that $n$ gives the number of  wave-function nodes), not $n=1$, see for instance Eq.~(\ref{levels_energy}). Keeping this tradition, we use $n_{x,y}=0,1,...$ for numbering the states of our system}. So the states $(0,n)$ are the bound atomic states. For $n \gg 1$ these are the Rydberg states. The state $(1,1)$ is the first doubly-excited state.  Using the modified P\"oschl-Teller potential (see Appendix) we can estimate the energy of this state in zero approximation neglecting the electron-electron interaction. Within this approximation the energy of the state is twice the energy $E_1$ given by eq.~(\ref{levels_energy}); this gives $2 \times -18.6$eV$=-37.2$eV, which is in a reasonable agreement with the level energy of $-34.0$ eV found numerically solving the TDSE. 

\section{Spatial distribution of the electronic density in the AIS}

The wave-functions for several states found via numerical TDSE solution are shown in Fig.~\ref{Wave-functions}, and details of the corresponding calculations are presented in Appendix, section~\ref{spatial}

Let us discuss the properties of the wave-functions shown in Fig.~\ref{Wave-functions}. First, the potential and the wave-functions are symmetric with respect to the line $y=x$. This is a result of the indistinguishability of the electrons. Note that we assume that the two electrons {\it can} find themselves in the same point; so we consider the parahelium. Second, in the figure one can see that the ground (0,0) and dark AIS (1,1) states are even, because the wave-functions do not change the sign after the inversion of the space coordinates $x \rightarrow -x$, $y \rightarrow -y$, and the first excited state (0,1) is odd (see the second and the third columns in Fig. ~\ref{Wave-functions}). Third, the AIS wave-function consists of the part which is well-localized near the origin, and the running waves which 'leak out' from the atom along the valleys ($x=0$ and $y=0$) of the potential (see Fig.~\ref{Potential3D}).

\begin{figure*}[]
\centering
\includegraphics[width=0.3\linewidth]{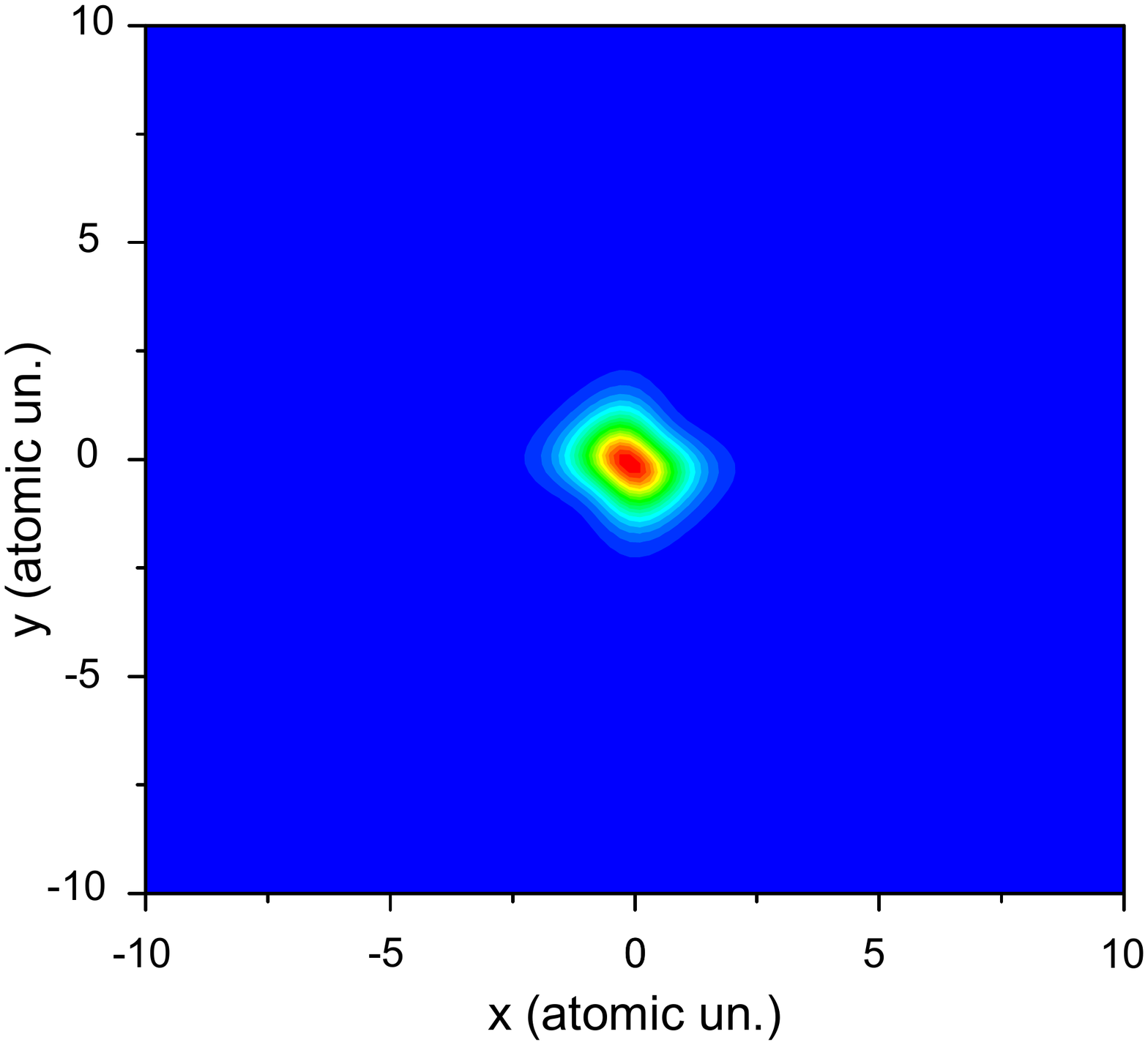}
\includegraphics[width=0.3\linewidth]{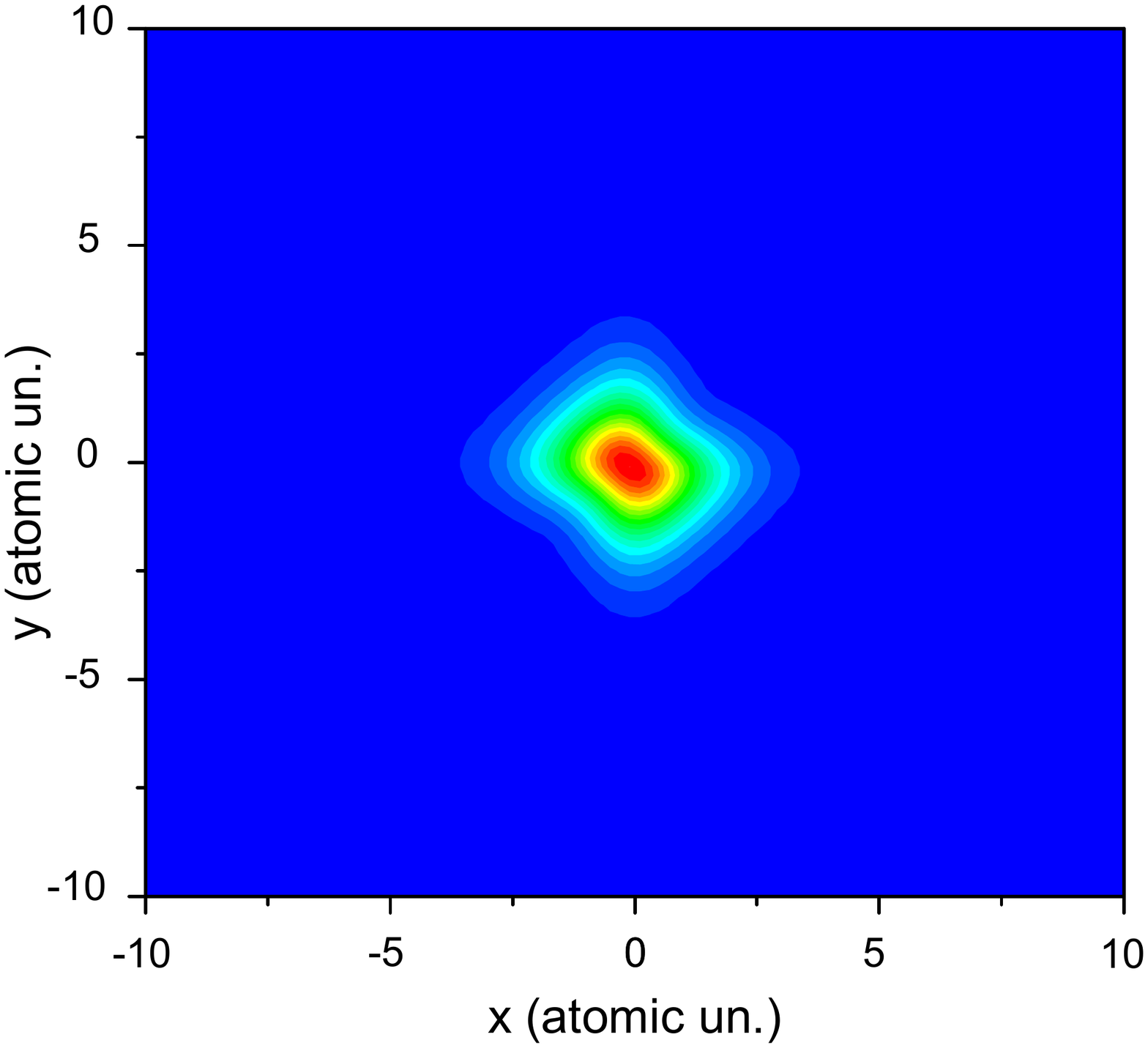}
\includegraphics[width=0.3\linewidth]{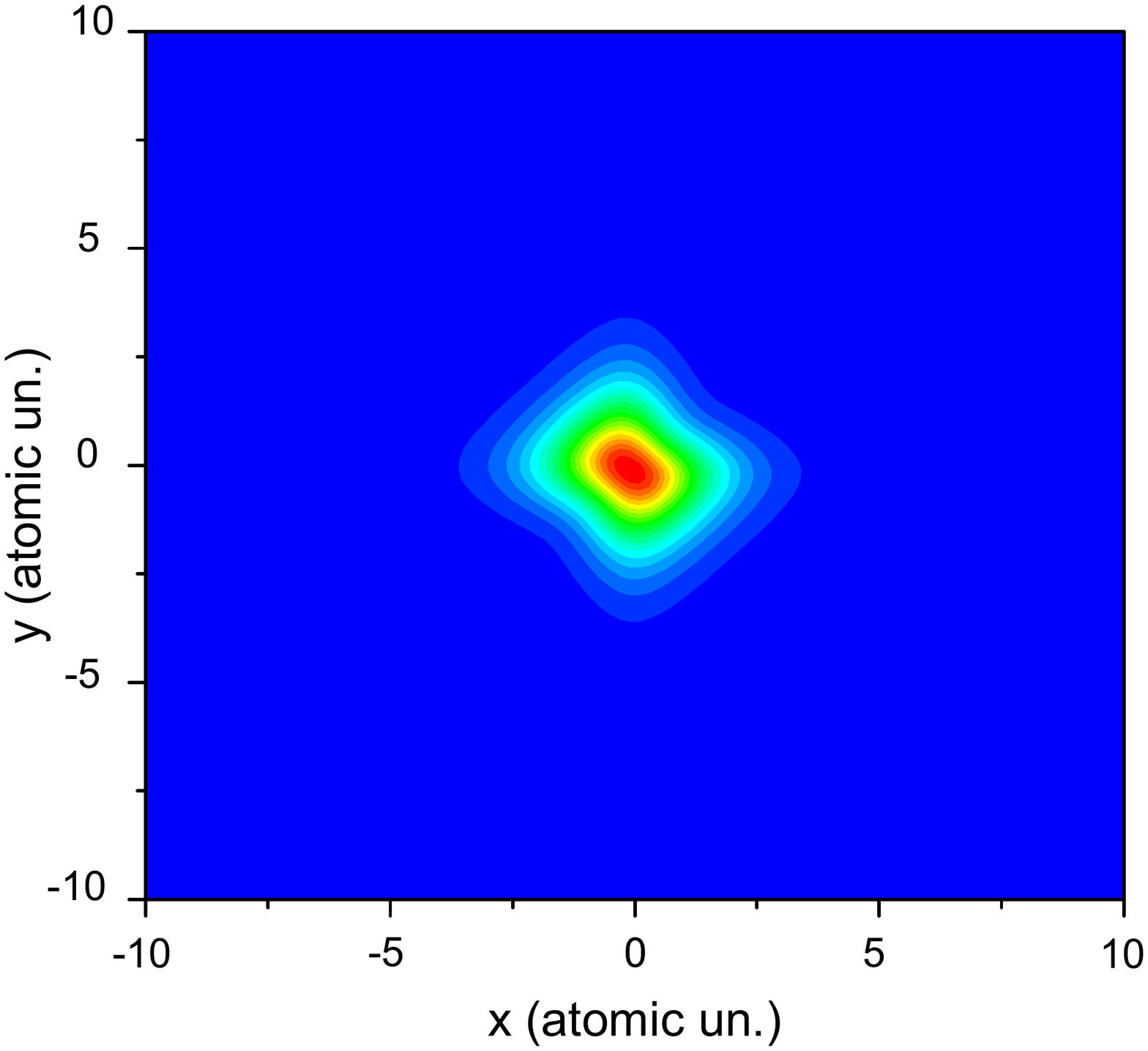}
\includegraphics[width=0.3\linewidth]{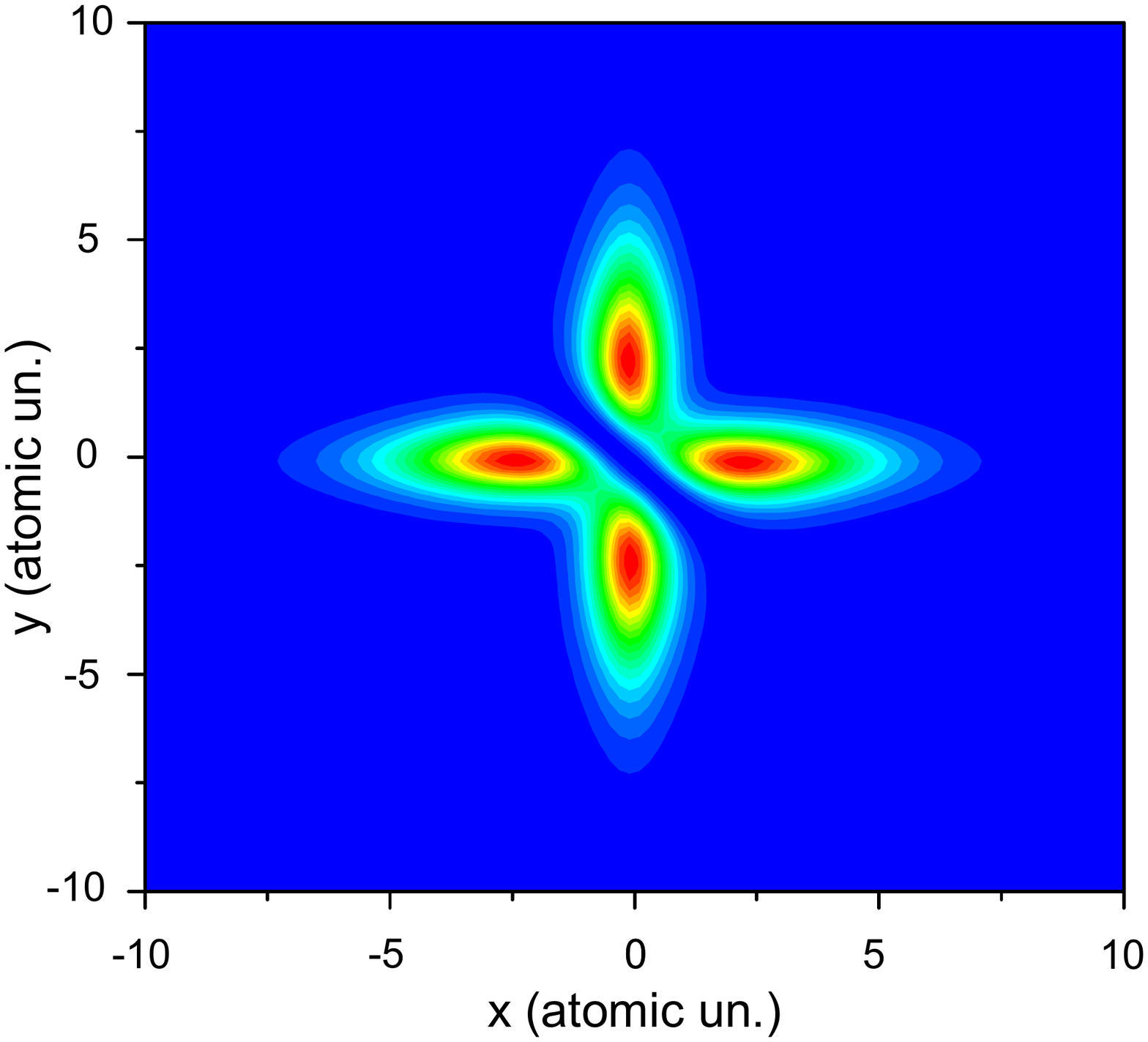}
\includegraphics[width=0.3\linewidth]{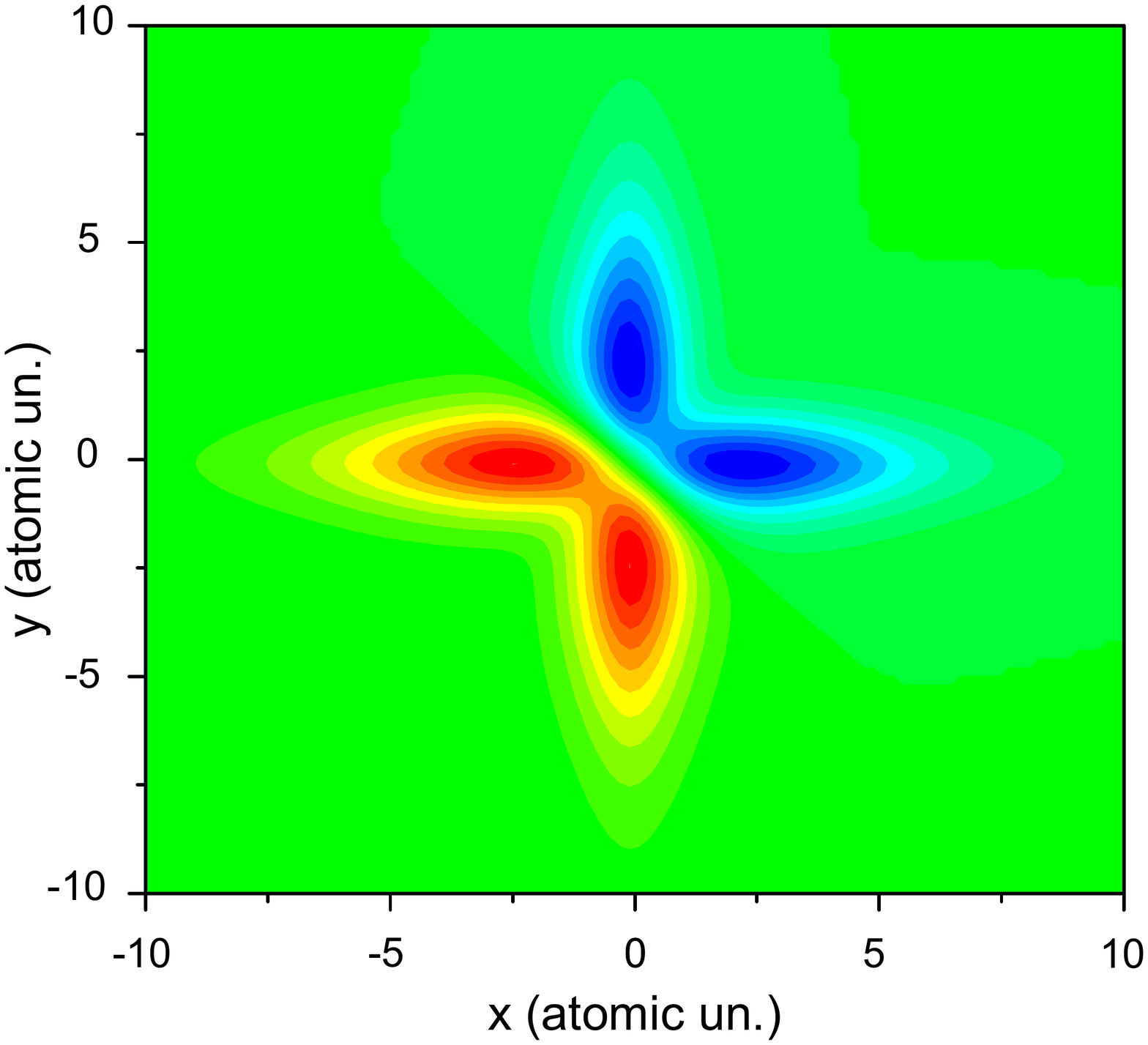}
\includegraphics[width=0.3\linewidth]{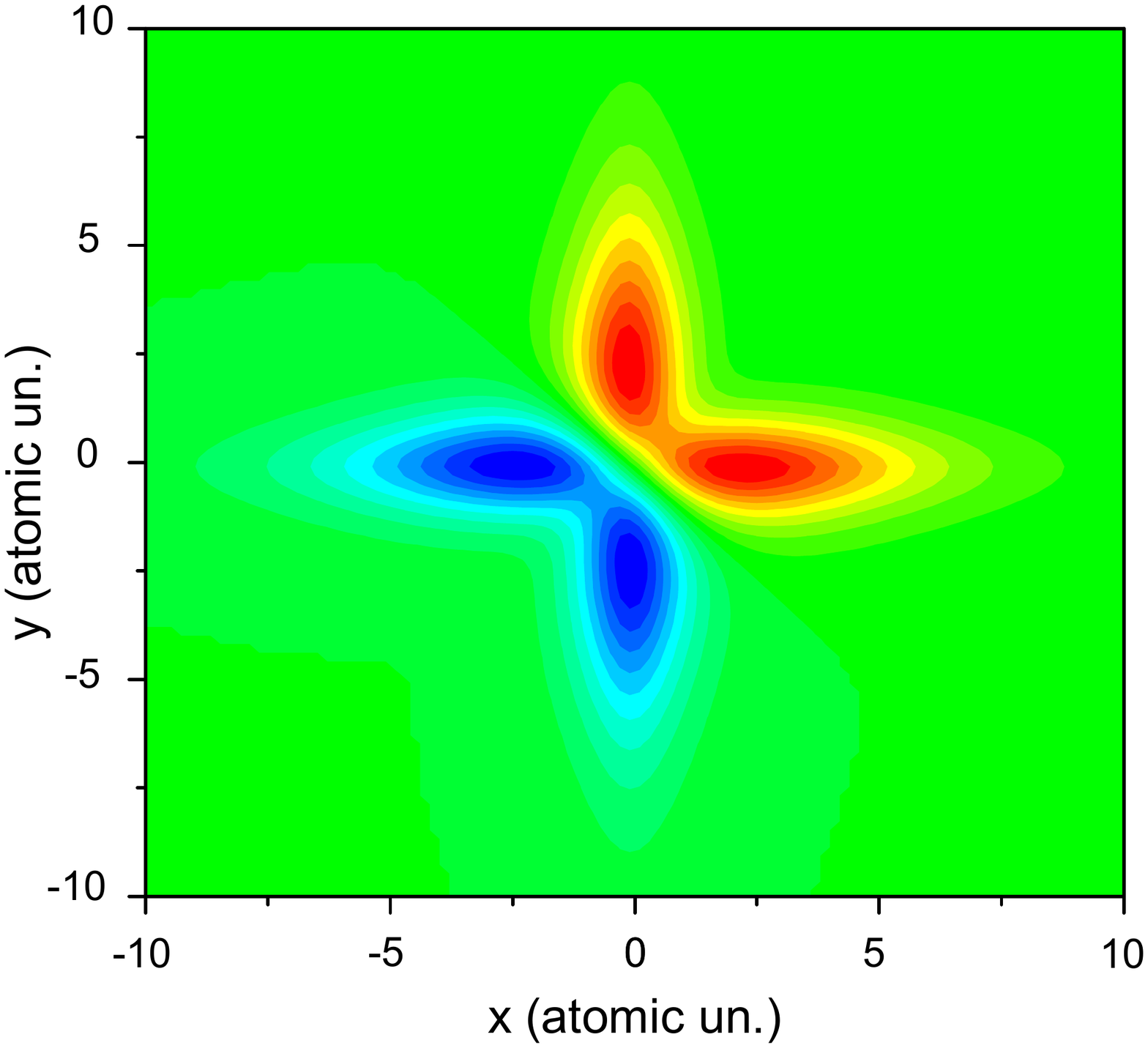}
\includegraphics[width=0.3\linewidth]{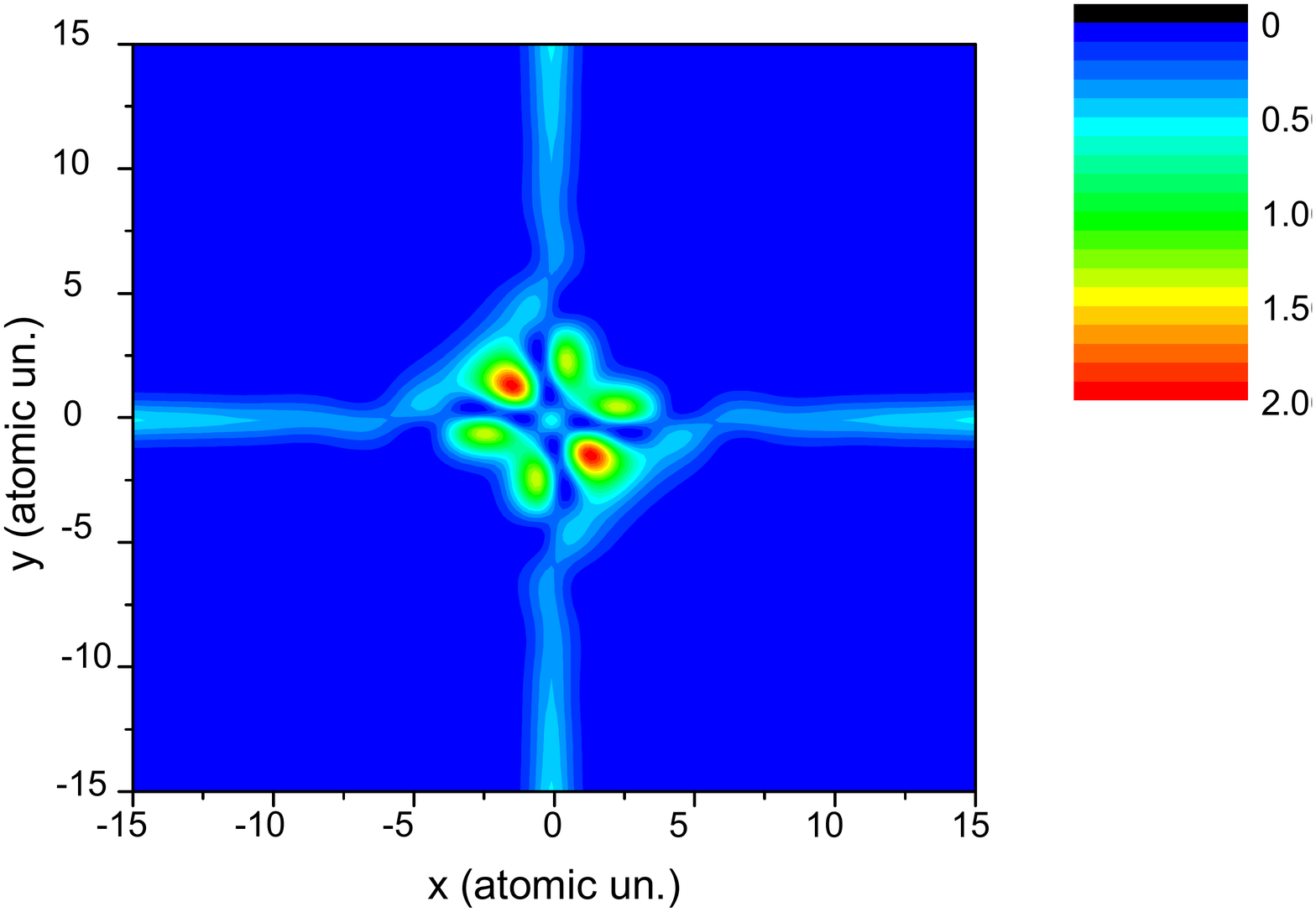}
\includegraphics[width=0.3\linewidth]{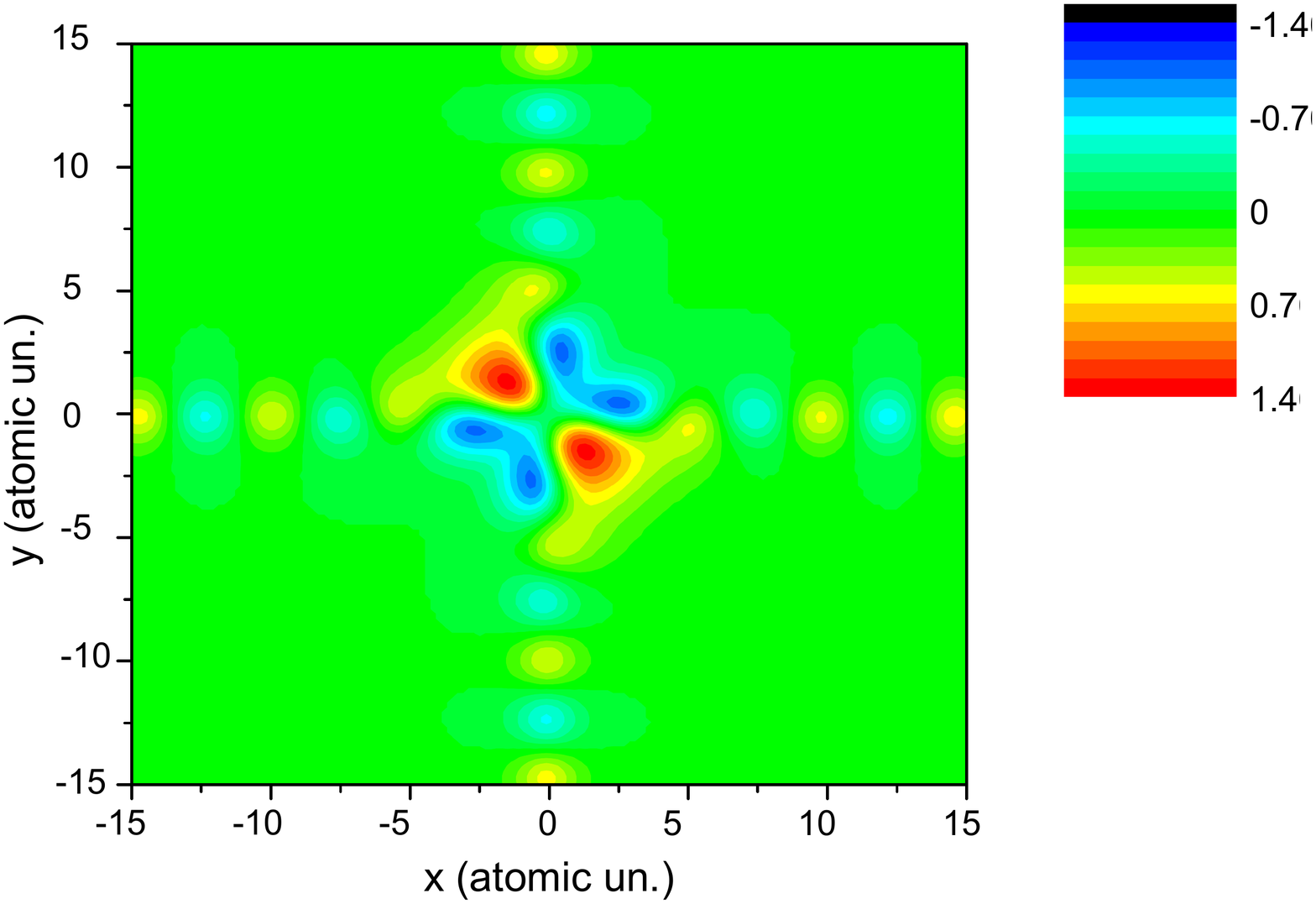}
\includegraphics[width=0.3\linewidth]{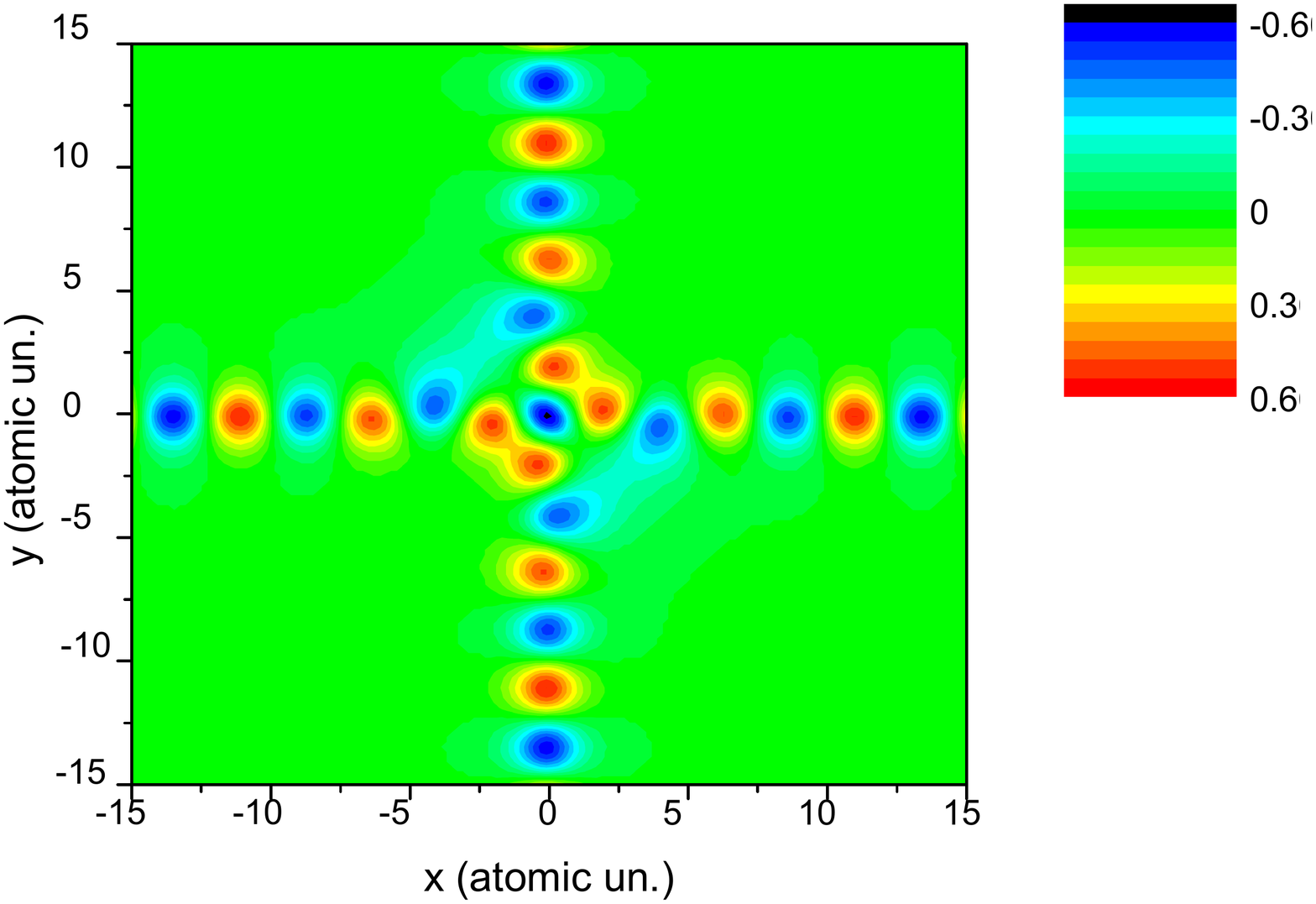}
\caption{Wave-functions of several states in 1D helium atom. Absolute value squared (left column), real (middle column} and imaginary (right column) part of the wave functions of ground (0,0) (upper row), first excited (0,1) (middle row) and lowest dark AIS (1,1) (lower row) states. 
\label{Wave-functions}
\end{figure*}

\section{Resonances with AI states in one- and two-photon ionization}
\begin{figure}[]
\centering
\includegraphics[width=0.95\linewidth]{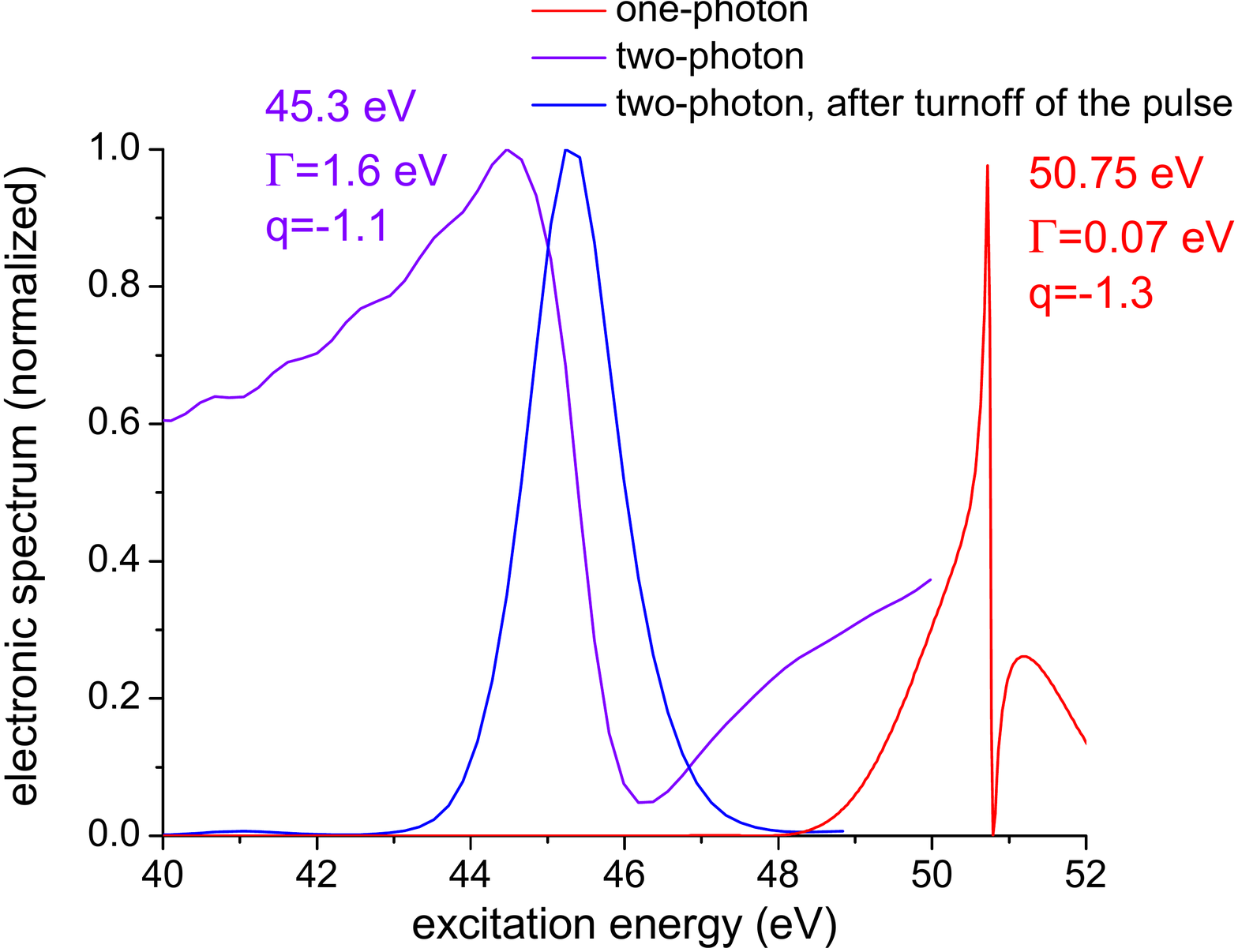}
\caption{Normalized photoelectron spectra calculated for the case of irradiation of the atom by one- (red curve) and two-color (blue and violet curves) XUV field, see the text for more details. The two-photon resonance leads to a Fano feature in the total spectrum (violet) due to the interference of the resonant and the non-resonant terms, whereas the resonant term can be separated calculating the spectrum over the time interval {\it after} the irradiation (blue). The central frequencies, the FWHM and the Fano parameter of the peaks are shown in the graph.}
\label{Cross-sections}
\end{figure}

In this section we study the resonant features appearing in the photoelectronic spectrum due to bright and dark AI states. To calculate the photoelectronic spectrum we solve numerically the TDSE for the atom irradiated with an XUV field, thus finding the wave-function as a function of space and time $\psi (x,y,t)$. Then we calculate\footnote{This operation requires storage of a big data in computer memory; to make it less demanding, we store the wave function not at {\it every} step of numerical integration $dt=0.05$, but at every 8-th step. Correspondingly, we find only the lowest-energy part of the photoelectronic spectrum. This is definitely sufficient here, because the inverse step of the numerical integration is much higher than the reasonable electronic energies.} its spectrum $\psi (x,y,\omega)$ and the energy distribution for the part of the wave-function, where at least one electron is free:
\begin{equation}
\widetilde W(\omega)= \int_{x^2+y^2>r^2} dx dy |\psi (x,y,\omega)|^2.
\label{photoelectron_spectrum}
\end{equation}
We used $r=30$ and checked that results are not sensitive to this parameter provided that $r > 20$. In the latter equation $\omega$ is the binding energy (negative when the system is in atomic bound states or in ionic bound states + free electron). It is more convenient to deal with the excitation energy $\omega-E_0$, where $E_0=-79.3$eV is the atomic ground state energy.  In the excitation energy range between the first ionization energy (24.9 eV) and the first excited state of the ion (52.4 eV) the photoionization leads to the appearance of a free electron and an ion in the ground state. So the energy of the ion is defined, thus the spectrum (\ref{photoelectron_spectrum}) directly corresponds to the photoelectron spectrum. Fig.~\ref{Cross-sections} presents the spectrum as a function of the excitation energy. 

The {\it bright} AI states lead to resonant features in the spectrum of the electrons detached due to {\it one}-photon ionization. To find this spectrum we solve the TDSE for the atom irradiated by a short XUV pulse. The pulse intensity is $3.5 \times 10^{12}$ W/cm$^2$ (the field amplitude is $0.1$ atomic unit), its duration is 100 as. The AIS leads to a Fano~\cite{Fano1961} feature in the photoelectron spectrum. Note that a similar feature was found in the 1D helium photoelectron spectrum in Ref.~\cite{Zhao2012}. We have found the width and the Fano parameter of the line, which is shown in the graph. 




To find the resonances with the {\it dark} AIS, we solve the TDSE for an atom irradiated by {\it two}-color XUV pulse. The carrier frequencies are 130 eV and 85 eV, the pulse duration is 100 as, the fields' amplitudes are 1.0 atomic units. These frequencies of the fields are chosen to have two-photon response in the studied spectral range (due to Raman-type transition 130-85=45 eV), and no one-photon response in this range. The short duration of the pulse provides the wide spectral range covered by the two-photon response (more precise, the spectrum of the field squared has a wide peak near 45 eV, and the FWHM of this peak is 27 eV). The TDSE is solved over the time interval of 20 fs thus over the irradiation time plus a long time after the pulse. For the attosecond pulse used here the latter time period is important for the correct photoelectron spectrum calculation because the AI states continue decaying after the pulse has already passed. Calculating the spectrum over the whole time we find the spectrum with the Fano feature, see the violet curve in Fig.~\ref{Cross-sections}. This feature appears due to the interference of the non-resonant and resonant contributions to the two-photon ionization; the latter contribution is due to the dark AIS. Moreover, calculating the spectrum for the time interval where the field is off, we find only the resonant contribution, see the blue line.  

Note that the authors of  Ref.~\cite{Haan1994} have studied the parameters of the Fano resonance appearing due to the single-photon ionization from the first atomic excited state to the dark AIS. The width of the resonance is very close to that found in our simulations (hence the parameters of the 1D He potential in Ref.~\cite{Haan1994} are $a=b=1$). The Fano parameter is also negative but it is much higher, thus the line shapes of the single-photon transition from the first excited state to the (1,1) AIS and of the two-photon transition from the ground state to the (1,1) AIS differ essentially.  

\section{Resonances with AI states in HHG}

In this section we study the role of the resonance with the AI states in HHG simulating the XUV spectrum emitted by our model atom in an external laser field. The peak pulse intensity is $8 \times 10^{14}$ W/cm$^2$,  the fundamental frequency $\omega_l$ is about 3 eV (the wavelength is about 400 nm, the cut-off is approximately at H19-H21, the Keldysh parameter is $\gamma \approx 0.95$); it is tuned near the 14th order resonance with the transition from the ground to the dark (1,1) AIS and near the 17th order resonance with the bright (1,2) AIS. The laser pulse intensity increases during 4 cycles, then it is constant during 8 cycles, than decreases down to zero during 4 cycles; the slopes of the pulse have $\sin^2$ shape. 

\begin{figure}[]
\centering
\includegraphics[width=0.95\linewidth]{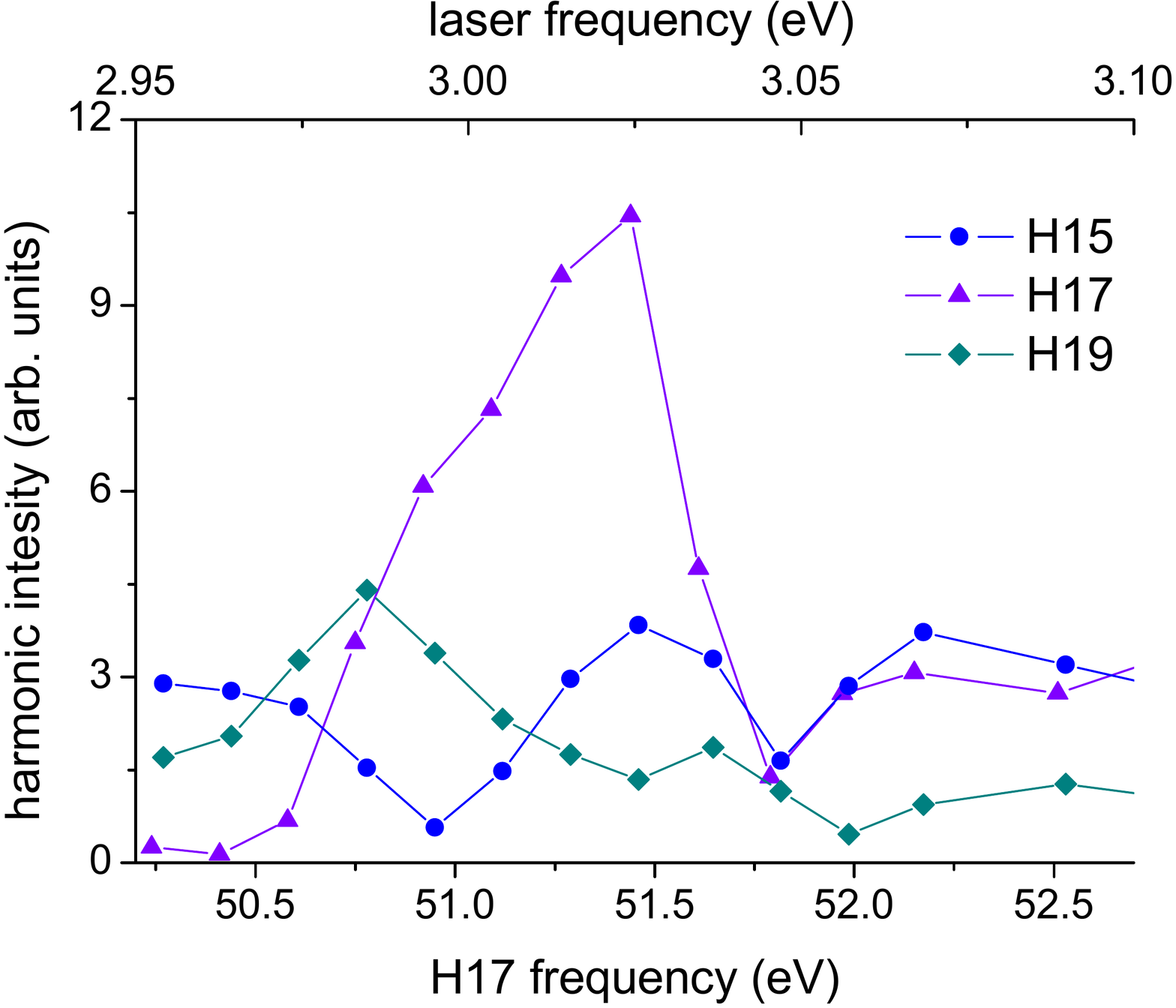}

\includegraphics[width=0.95\linewidth]{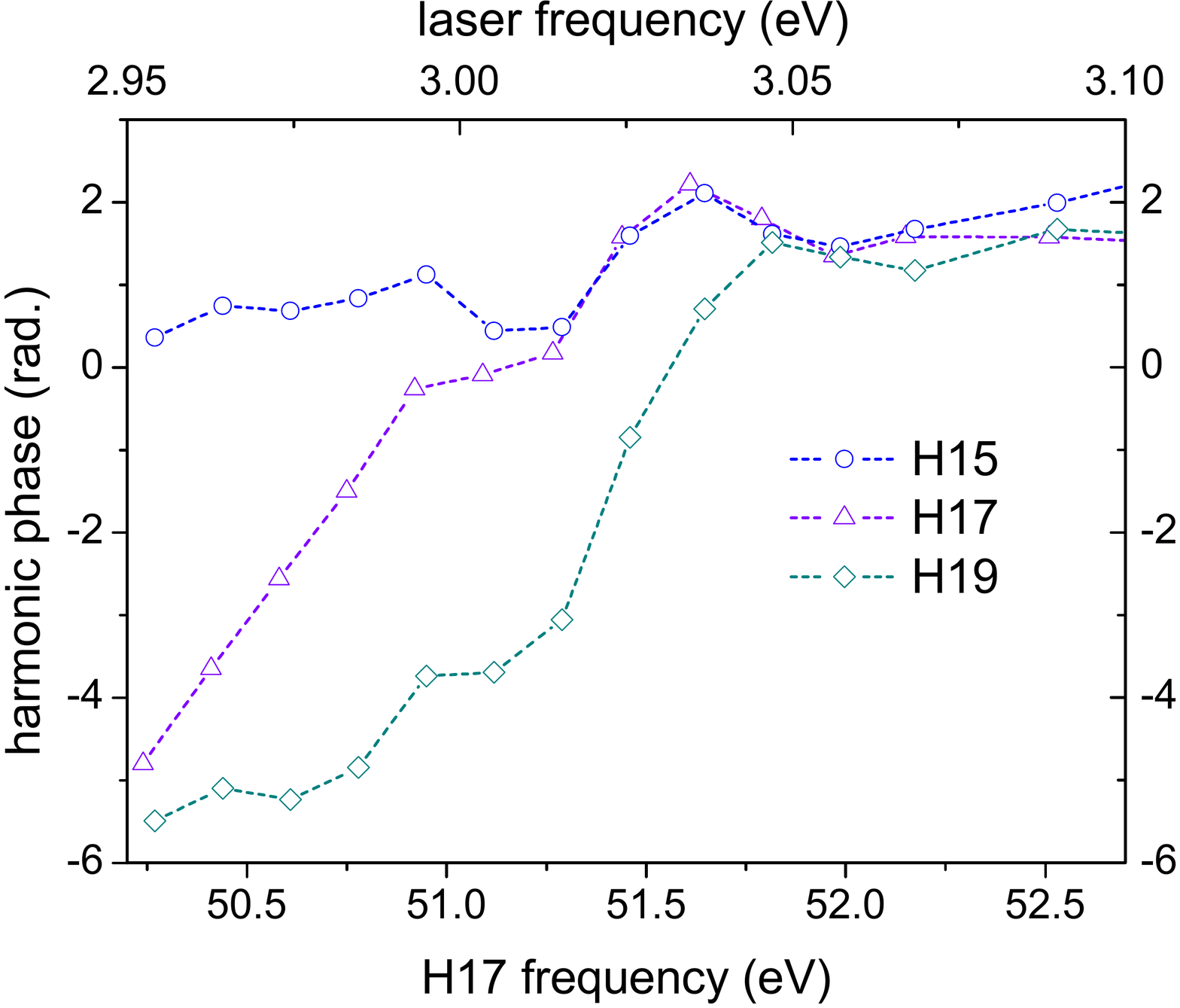}
\caption{Intensity (a) and phase (b) of harmonics 15 (blue circles), 17 (violet triangles), and 19 (dark cyan diamonds) in the vicinity of the 17-photon resonance with the bright AIS (1,2).}
\label{Resonance_bright}
\end{figure}

In Fig.~\ref{Resonance_bright} we show the intensities and phases of H15, H17, and H19 as functions of laser frequency (the horizontal axis above the graph) in the vicinity of the resonance with the bright AIS. The H17 frequency is shown for reference in the axis below the graph.  We can see the pronounced enhancement of the resonant harmonic intensity. The "enhancement line" (enhancement as a function of the laser frequency) is shifted and broadened with respect to the resonant line in the one-photon photoelectron spectrum shown in Fig.~\ref{Cross-sections}. This can be attributed to the line modification by the laser field (to the Stark shift of the levels and photoionization of the AIS). Moreover, there is even qualitative difference between the shapes of the lines in the photoelectronic spectrum and harmonic enhancement: in the spectrum the dip is at the high-energy side of the maximum, in the enhancement line it is at the low-energy side. This agrees with the experimental studies in helium: in the photo-absorption spectrum~\cite{Chan_helium} the dip is at the high-energy side, and in the XUV spectrum~\cite{Chang2008} it is at the low-energy one. Similar difference was discussed in~\cite{Strelkov2014}. 

One can also see a pronounced modification of the harmonic phase by the resonance; such distortion was measured in~\cite{Haessler2013}. Also there is some enhancement of the harmonic above the resonance (H19) and almost no enhancement of the harmonic below it (H15). However, the phases of all the three harmonics are affected by the resonance. The total variation of the laser frequency in Fig.~\ref{Resonance_bright} is small (approx 5\%), so the non-resonant contribution to the generation of a harmonic should not vary much; the resonant one vanishes at the edges of the considered frequency range. Thus, the total phase variation over this range for a harmonic is approximately either $0$ or $2 \pi$. 

The HHG modification by dark AIS has been studied much less than that for the bright AIS. Very recently the resonance of an even number of laser photons with the dark AIS was considered~\cite{Fareed2022} and an enhancement of the harmonic with an order differing by unity from this number was observed. This phenomenon can be understood as a resonance between the harmonic and the dark AIS, dressed by a laser photon. 

\begin{figure}[]
\centering
\includegraphics[width=0.95\linewidth]{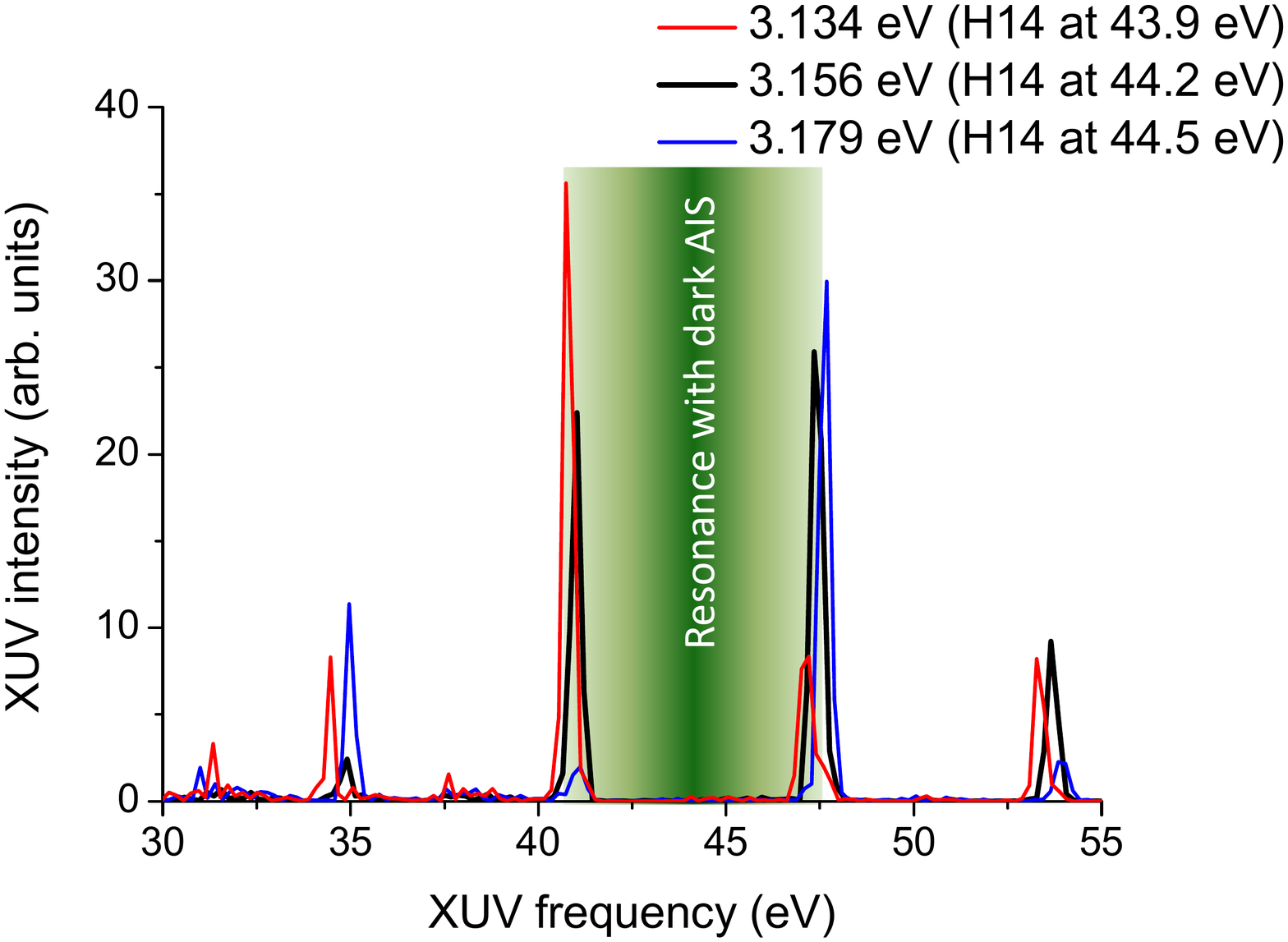}
\caption{HHG spectra in the vicinity of the 14-photon resonance with the dark AIS (1,1). The fundamental frequency and the frequency of the 14 harmonic is shown in the graph.}
\label{HHG}
\end{figure}

Our simulations presented in  Fig.~\ref{HHG} show that in the case of the 14th order resonance with the dark AIS the two neighbouring harmonics (i.e. H13 and H15) are enhanced by the resonance.
Note that there are no one-photon resonances that can cause the HHG enhancement in the considered spectral range (see Fig.\ref{Cross-sections}).

\begin{figure}[]
\centering
\includegraphics[width=0.95\linewidth]{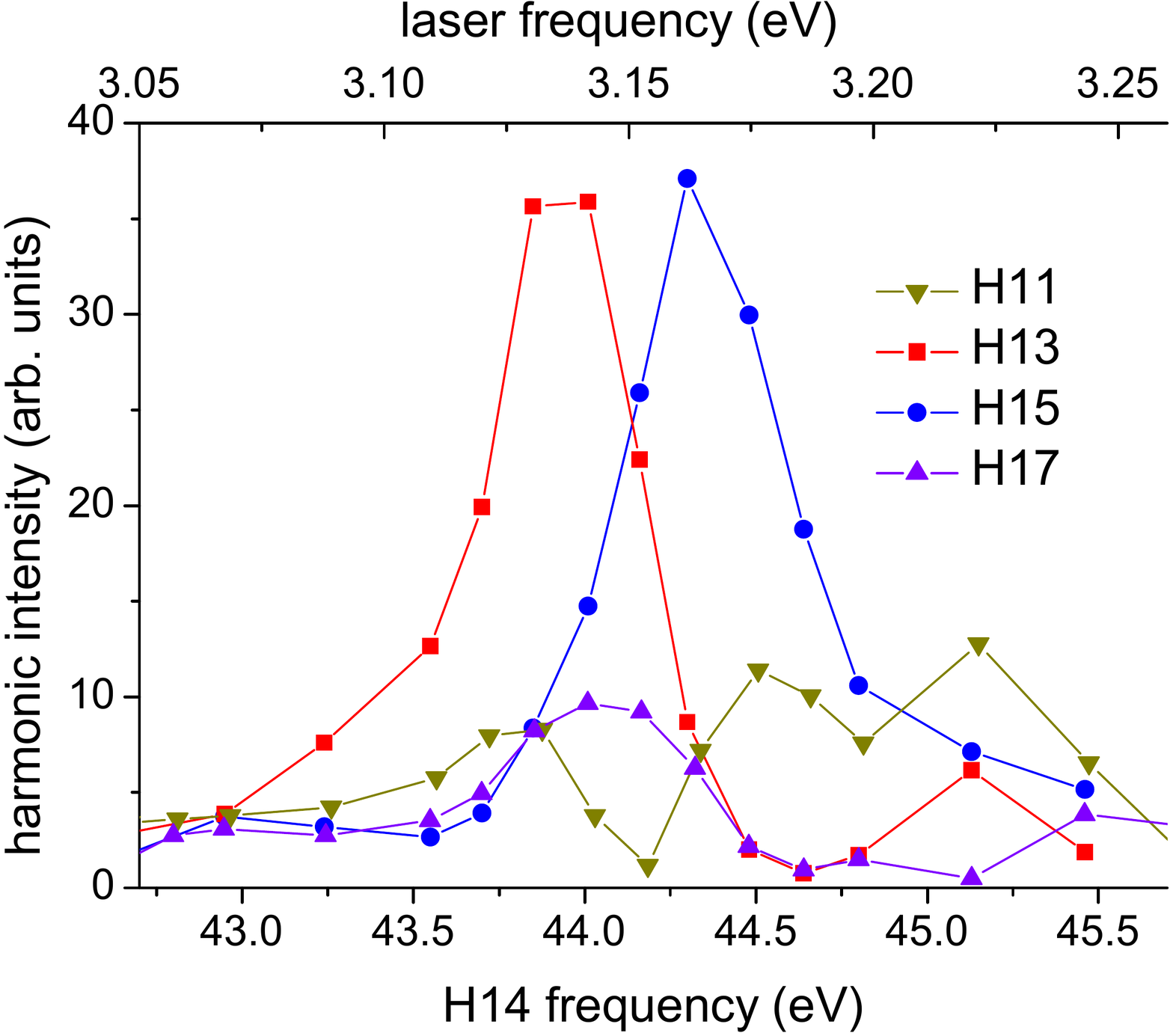}

\includegraphics[width=0.95\linewidth]{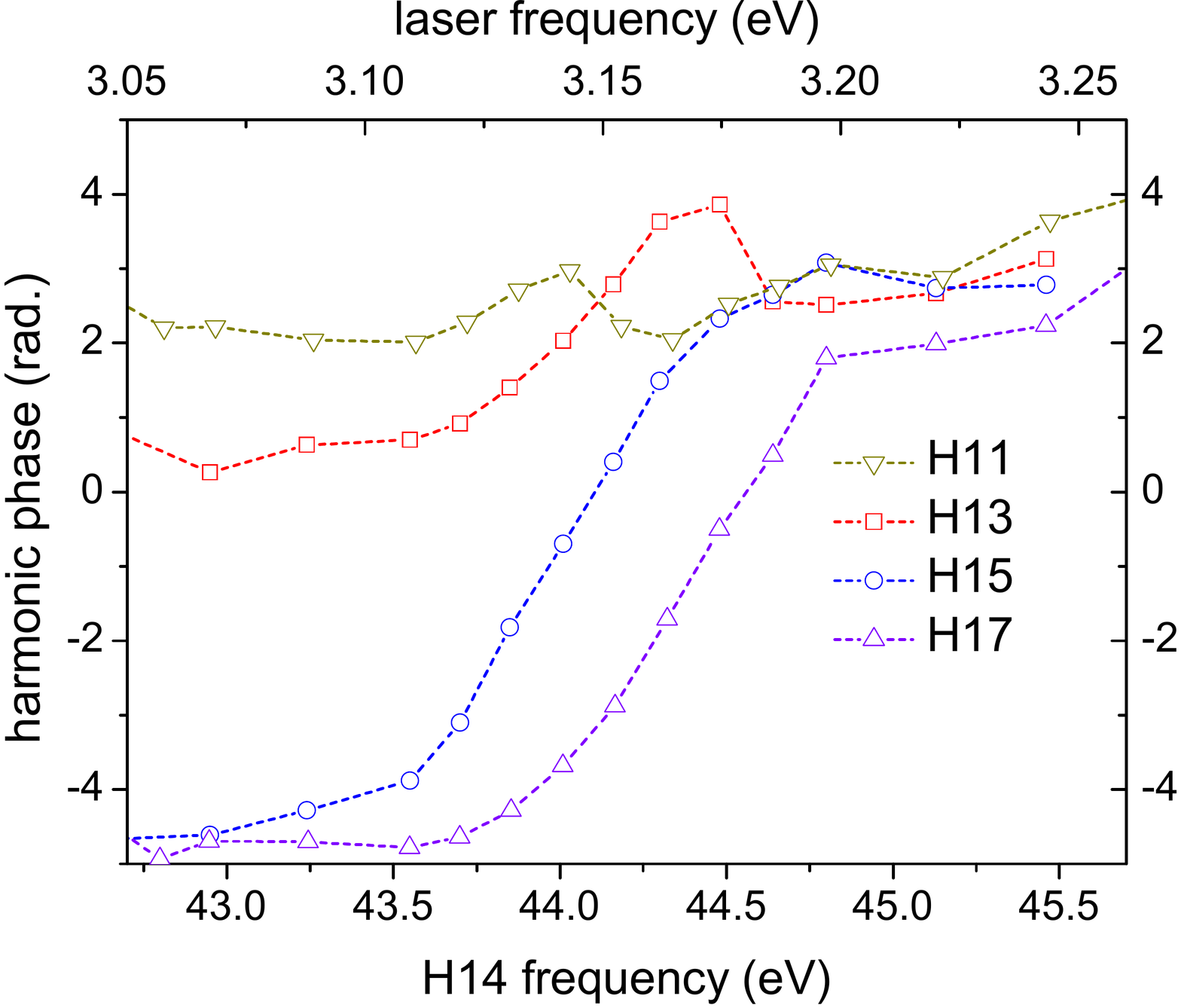}
\caption{Intensity (a) and phase (b) of the harmonics 11 (yellow triangles), 13 (red squares), 15 (blue circles), and 17 (violet triangles) in the vicinity of the 14-photon resonance with the dark AIS (1,1).}
\label{Resonance_dark}
\end{figure}

To study this enhancement in more detail we present in Fig.~\ref{Resonance_dark} the intensity of H13 and H15 as functions of the fundamental frequency in the vicinity of the 14-photon resonance with the dark AIS. Comparing this figure with Fig.~\ref{Resonance_bright} we can see that the enhancements caused by the resonances with dark and bright AIS are comparable in the considered conditions. The width of the enhancement lines for H13 and H15 are close to the one in the two-photon photoelectron spectrum in Fig~\ref{Cross-sections}; the center of the HHG enhancement line is slightly shifted due to the Stark shift of the levels in the laser field. Counterintuitively, the enhancement lines for H13 (red) and H15 (blue) are different: the peak enhancement is achieved for different laser frequencies; moreover, zero enhancement for H13 near 44.6eV does not correspond to any specific feature of H15 at this frequency.

The difference of the enhancement lines for H13 and H15 can be understood taking into account that XUV emission time is different under resonant and non-resonant conditions: one can assume that, similar to the HHG enhancement with the bright AIS, the emission of the resonant XUV is delayed (with respect to the non-resonant one) by the AIS lifetime~\cite{Tudorovskaya2011, Strelkov2016}.  In more detail, the difference of the harmonic enhancement lines can be explained as following.

Similarly to the Fano line in the photoionization cross-section, the harmonic enhancement line originates from the interference of the resonant and non-resonant terms~\cite{Strelkov2014} in the harmonic amplitude. The phase of the resonant term varies strongly near the resonance as a function of the detuning, whereas the phase of the non-resonant term does not. Let us denote the phase difference of these terms for the $q$th harmonic as 
\begin{equation}
\delta \varphi_q= \varphi_q^r- \varphi_q^{nr}   
\label{delta_phi}
\end{equation}.


The difference of H15 and H13 phases $\Delta \varphi=\varphi_{15}-\varphi_{13}$ defines the emission time~\cite{Mairesse2003} of the attosecond pulse consisting of H15 and H13:  $t_e=\Delta \varphi /(2 \omega_l)$. (Note that $\Delta \varphi$  should not be confused with $\delta \varphi_q$: the former gives the phase difference between the neighbouring harmonics, whereas $\delta \varphi_q$ describes the interference of the resonant and non-resonant terms of the same harmonic)

Far from the resonance the non-resonant term dominates: $\varphi_q \approx \varphi_q^{nr} $, so the emission time is defined by this term:
\begin{equation}
\Delta \varphi^{nr}=\varphi_{15}^{nr}-\varphi_{13}^{nr}=2 t_e^{nr} \omega_l,
\label{Delta_phi_r}
\end{equation}
Note that the phases of the non-resonant terms do not vary much within the considered small detuning interval of the fundamental, in particular, this equation is valid near the resonance as well.

Near the resonance the resonant term dominates: $\varphi_q \approx \varphi_q^{r} $, so the emission time is defined by this term. This emission time as we mentioned above is $t_e^{r}= t_e^{nr}+ \tau$, where $\tau$ is the AIS lifetime. So near the resonance  
\begin{equation}
    \Delta \varphi^{r}= \varphi_{15}^{r}-\varphi_{13}^{r}=2  (t_e^{nr}+ \tau) \omega_l.
     \label{Delta_phi_nr}
\end{equation}

From Eqs.~(\ref{delta_phi}), (\ref{Delta_phi_r}), and (\ref{Delta_phi_nr}) we conclude that near the resonance
$\delta \varphi_{15}-\delta \varphi_{13}= 2 \tau \omega_l$. So the enhancement line shape defined by the interference of resonant and non-resonant contributions to the harmonic emission is different for H13 and H15.  

From the harmonic phases shown in Fig.~\ref{Resonance_dark} we find that $\Delta \varphi^{nr} \approx 0$ and $\Delta \varphi^{r} \approx 2.8$~rad. From  Eqs.~(\ref{Delta_phi_r}) and (\ref{Delta_phi_nr}) we find $\tau=300$~as. This agrees with the fact that this delay is comparable but less than the inverse FWHM of the resonant term in the two-photon ionization cross-section (blue line in Fig.~\ref{Cross-sections}), namely $1/\Gamma=470$~as.

Note that in Ref.~\cite{Haessler2013} the resonant phase was measured under some detuning from the resonance because closer to the resonance the intensity difference of the resonant and non-resonant harmonic make RABBIT measurements problematic. For HHG enhanced by the dark AIS such measurements can be done near the center of the resonance, because the {\it two} harmonics are enhanced in a similar way. 

\begin{figure}[]
\centering
\includegraphics[width=1.0 \linewidth]{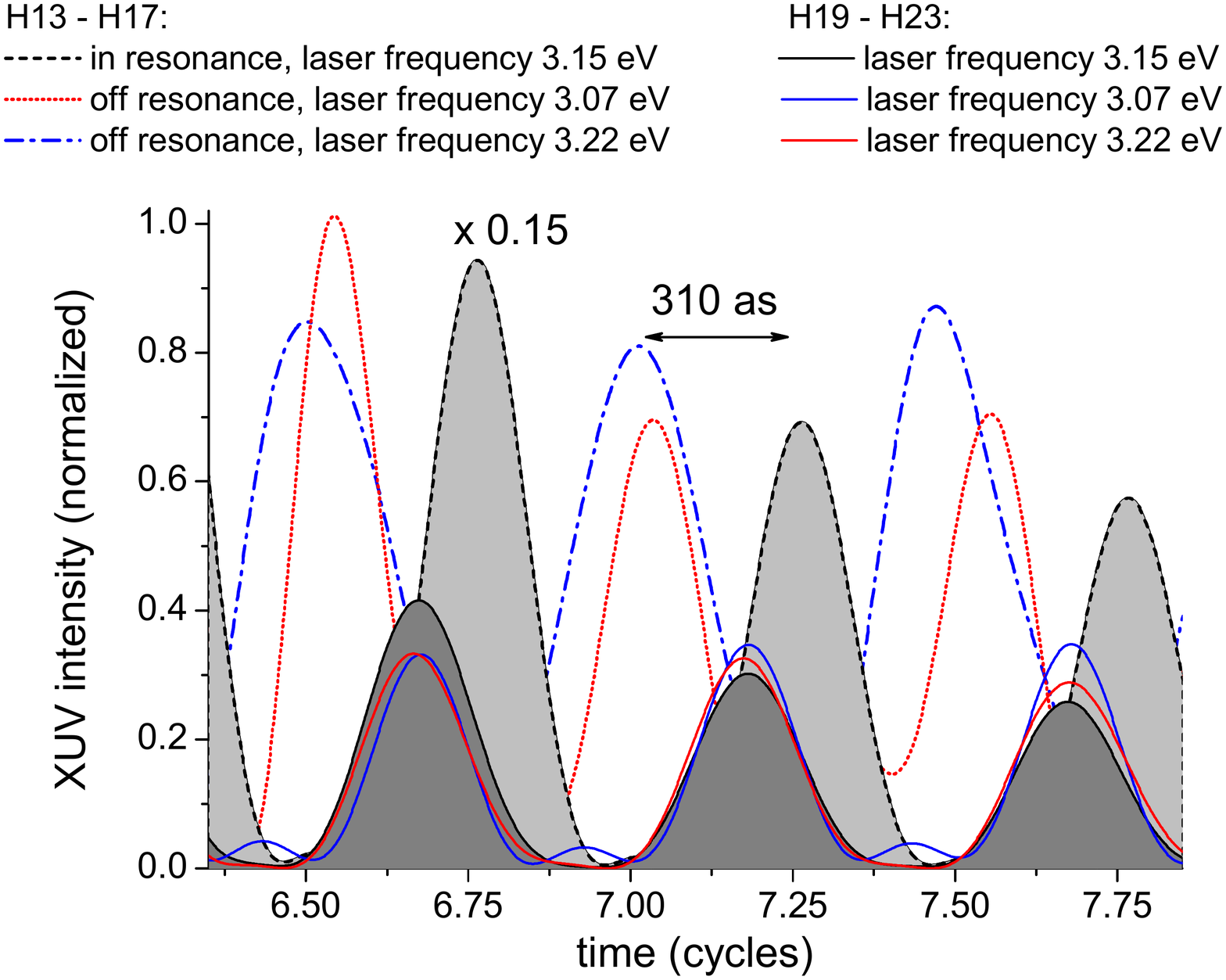}
\caption{Attosecond pulses produced by the harmonics near the resonance with the dark AIS (dashed, dotted, and dash-dotted lines) and by the cut-off harmonics (solid lines) for the three fundamental frequencies shown in the graph.}
\label{Attopulses}
\end{figure}

Fig.~\ref{Attopulses} shows the attosecond pulses obtained from the cut-off harmonics (H19-H23) and harmonics near the 14-photon resonance with dark AIS (H13-H17). Naturally, the emission time of the cut-off attosecond pulse does not depend on the detuning from the resonance. So this emission time gives a straightforward reference for the emission times of the attosecond pulses obtained from H13-H17. The latter pulses in the off-resonance conditions are emitted before the cut-off attosecond pulse. (This shows that the short quantum path dominates in HHG response, see~\cite{Mairesse2003}.) The H13-H17 pulses are emitted approximately at the same time for the above-resonance conditions (laser frequency 3.22eV) and the below-resonance ones (laser frequency 3.07 eV), so the resonance does not affect the attosecond pulses emission. In the resonant case the attosecond pulse is much more intense and emitted later than in the off-resonant cases, in agreement with the considerations presented above. The delay of $310$~as is close to the one found above from the harmonic phases.

\section{Conclusions}
In this paper we study the role of the dark and the bright AIS in resonant HHG using the 1D helium model allowing ab initio numerical TDSE simulations beyond the singe-electron approximation. 

Analytical model using a modified P\"oschl-Teller potential allows obtaining a reasonable estimate of the lowest double-excited state energy. We present the structure of the (lowest) levels of the studied system. The simplicity of the system allows straightforward graphical presentation of the wave-functions of different states, including the AI ones. 

We make detailed simulations of the interaction of the atom with electromagnetic field. The photoelectronic spectra demonstrate pronounced features due to resonances with AIS. In particular, the resonances lead to asymmetric Fano peaks in the spectrum of the photoelectrons appearing due to one- and two-photon ionization. 

We simulate the HHG enhancement via the resonance with the bright AIS and find a pronounced difference between the shapes of the above-mentioned Fano maximum in the photoelectronic spectrum and the harmonic enhancement line. We find an essential enhancement of the resonant harmonic, some enhancement of the harmonic above it and no enhancement of the harmonic below it.

Moreover, we simulate HHG under the conditions when the fundamental frequency is close to a multiphoton resonance of an even order with the dark AIS. We find an enhanced generation of the neighbouring harmonics. The shapes of the enhancement lines for the harmonics are different. The difference can be understood taking into account the harmonic phase properties defined by a delay between the emission of the non-resonant and resonant XUV; this delay is close to the AIS lifetime. Simultaneous enhancement of the two harmonics by the resonance with dark AIS makes feasible the experimental measurement of resonance-induced dephasing between them. Finally, our simulations show that the enhancement due to dark and bright AIS is comparable in the studied system.  


\section{Acknowledgments}
We are grateful to V.~Birulia for the TDSE code development, and to M. Singh, M. A. Fareed, T. Ozaki, A.I. Magunov, A. N. Grum-Grzhimailo for the discussions on the AI states properties and their role in HHG. We would like to thank N. Yu. Shubin for his helpful language assistance.
The study was funded by RSF (grant No 22-12-00389). 

\section*{Appendix}

\setcounter{section}{0}
\setcounter{figure}{0}
\setcounter{equation}{0}
\renewcommand{\thesection}{\Roman{section}}
\renewcommand{\thefigure}{A\arabic{figure}}
\renewcommand{\theequation}{A\arabic{equation}}

\subsection{1D hydrogen-like ion}
The 1D hydrogen atom or hydrogen-like ion with the Coulomb potential or the soft-Coulomb potential was studied in an number of papers~\cite{1D_H_1959, 1D_H_2006, 1D_H_2009, 1D_H_2016,1D_H_2018,1D_H_2021}. In contrast to the 3D system, the bound state of the 1D hydrogen cannot found analytically, either for the Coulomb, or for the soft-Coulomb potential.  Here we describe our numerical method and approximate analytical approach to find the eigenstates' energies and wave-functions.   

\begin{figure}[h]
\flushleft
a)

\includegraphics[width=0.9\linewidth]{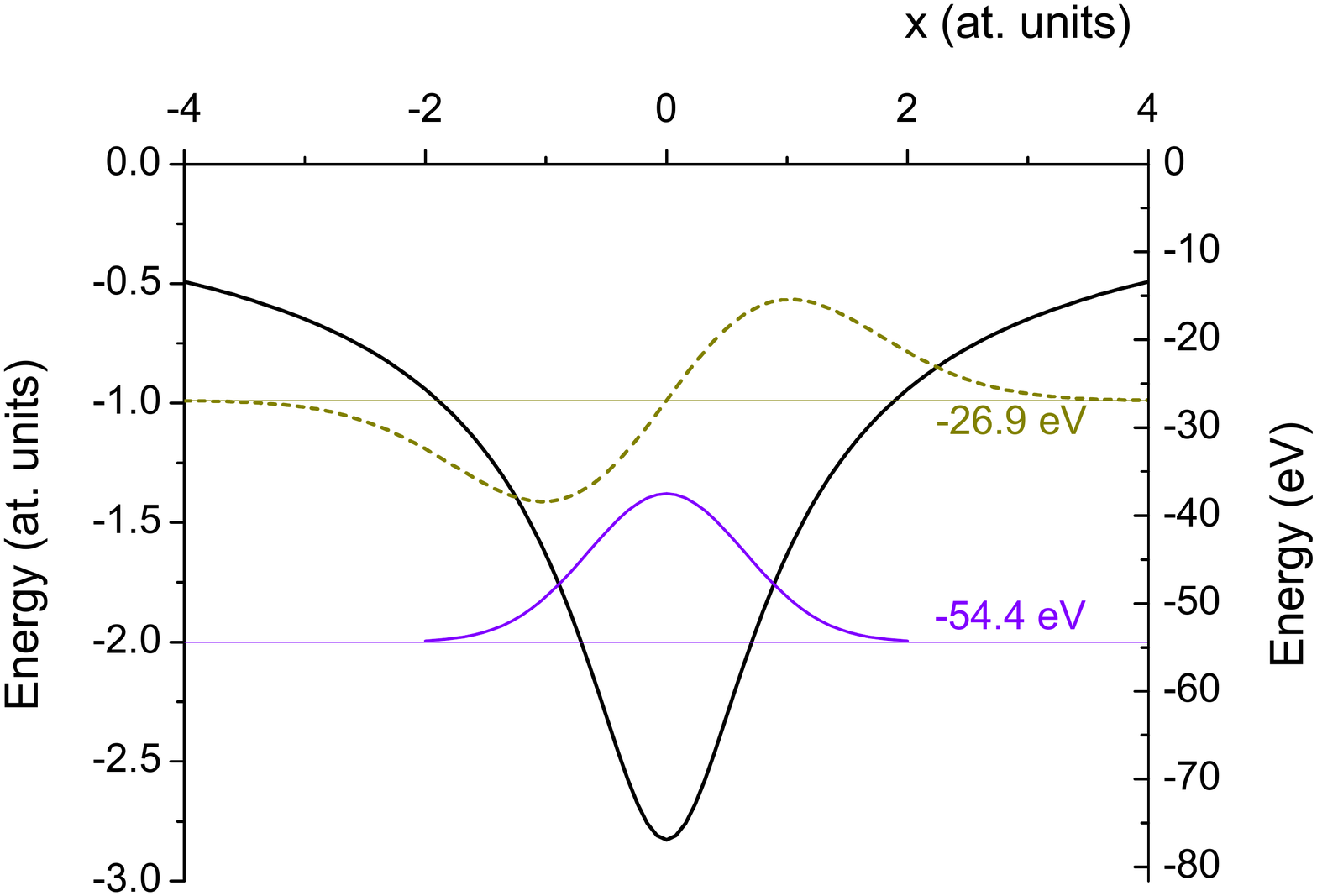}

\flushleft
b)

\includegraphics[width=0.9\linewidth]{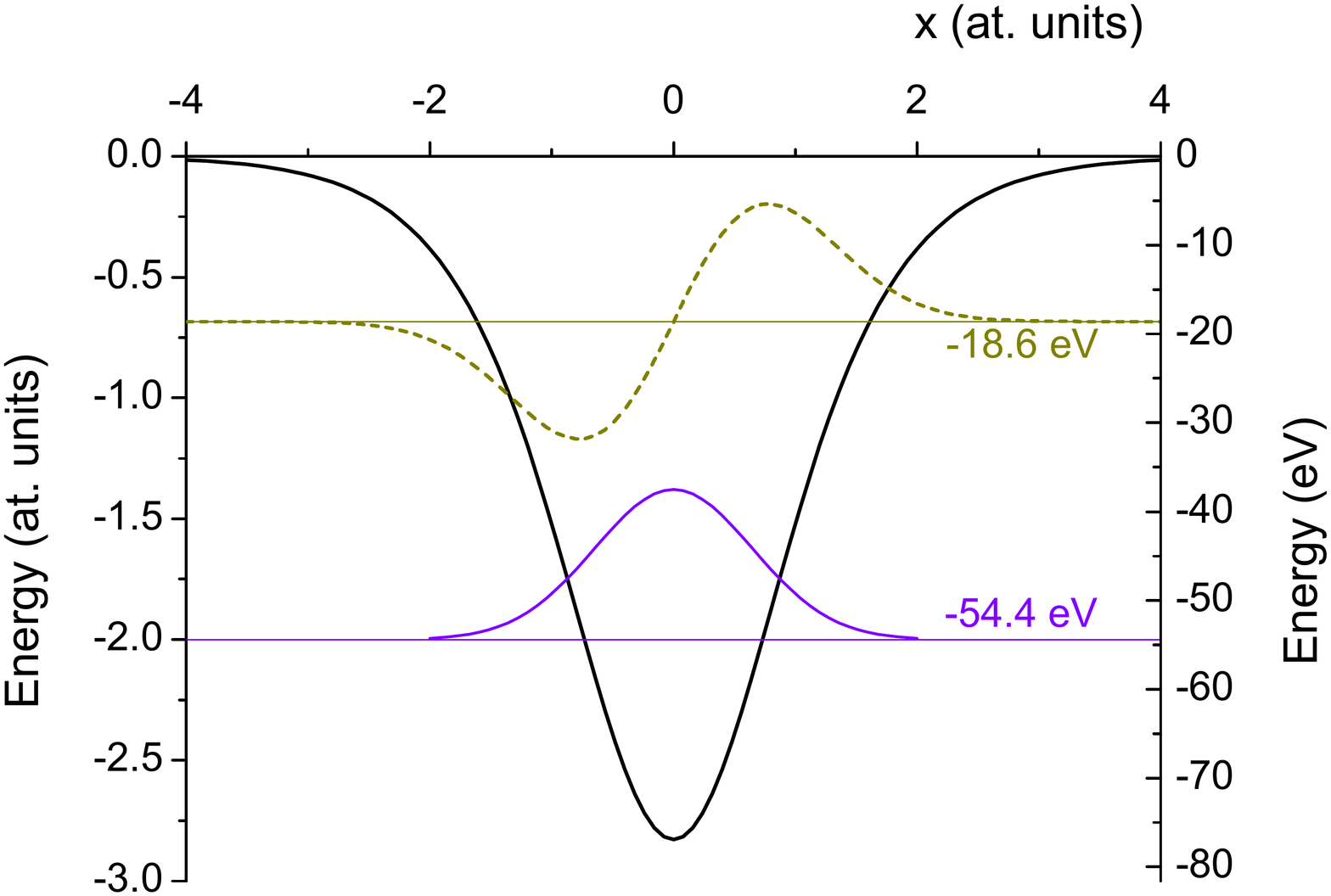}
\caption{Soft-Coulomb (a) and modified P\"oschl-Teller (b) potentials. Blue and red lines show the ground and the first excited state wave-functions, respectively.}
\label{Potentials}
\end{figure}

To study He$^+$ ion we remove the last term in Eq.~(\ref{eq_Potential}). Then the TDSE describes two non-interacting ions, and the interpretation of the TDSE solution results in terms of a single ion is straightforward. The ion's potential 
is 
\begin{equation}
V_{ion}(x)=\frac{-2}{\sqrt{x^2+a^2}}.
\label{ion_Potential}
\end{equation}
Under $a=1/\sqrt{2}$ the ground state energy of the model ion is equal to the actual ionization energy of He$^+$~\cite{Javanainen}. This soft-Coulomb potential is shown in Fig.~\ref{Potentials}a. 

Solving the TDSE we find the energies of the ground state and several first excited ones. The energy and wave-functions of the states are found as following. The initial wave-function of the $n$-th state ($n=0,1$...) is set as the corresponding wave-function of a harmonic oscillator; then the wave-function evolution is simulated via numerical TDSE solution. The part of the initial wave-function that is orthogonal to the bound eigenstates of the ion spreads in space and finally is absorbed by the absorbing boundary of the numerical box. So the part of the wave-function bound near the origin corresponds to the eigenstate of the potential~(\ref{ion_Potential}) (more precisely, it corresponds to a superposition of the eigenstates; however, choosing the initial wave-function with proper number of zeros guarantees that the weight of all eigenstates, except one, is very small). Numerical TDSE solution gives us the wave-function as a function of space and time $\psi (x,y,t)$. We calculate its spectrum $\psi (x,y,\omega)$ and the following energy distribution:
\begin{equation}
W(\omega)= \int dx dy |\psi (x,y,\omega)|^2,
\label{energy_distribution}
\end{equation}
(In contrast to Eq.~(\ref{photoelectron_spectrum}), here the integration is done over the whole numerical box.) The energy of the maximum in this distribution gives the energy of the found state; this negative energy is the state's binding energy, multiplied by $-1$. Other peaks in the energy distribution, if any, are much weaker; the ratio of the peaks to the main one gives the weights of the other eigenstates in the found state.

 In~\cite{1D_H_1959} the approximate equation for the ground states' energies of the truncated Coulomb potential was found. The accuracy of the approximation becomes better as the truncation parameter (roughly analogous to the parameter $a$) tends to zero. However, under $a=1/\sqrt{2}$ the accuracy for the ground and first excited state is insufficient. Much better analytical approach can be developed approximating the soft-Coulomb potential~(\ref{ion_Potential}) with the modified P\"oschl-Teller potential:    

\begin{equation}
V_{PT}(x)=-\frac{\alpha^2}{2} \quad \frac{\lambda (\lambda-1)}{\cosh^2(\alpha x)},
\label{PT_potential}    
\end{equation}
where $\alpha$ and $\lambda$ are parameters. The bound states' energies for this potential are~\cite{Flugge2}:

\begin{equation}
\begin{array}{l}
E_n=-\frac{\alpha^2}{2}(\lambda-1-n)^2,\\
n= 0,1,...\\
n \le \lambda-1,
\end{array}
\label{levels_energy}
\end{equation}
an even $n$ gives the energy of an even state, and an odd $n$ gives the energy of an odd one.  

The parameters $\alpha$ and $\lambda$ are chosen so that: 

- the minimum of the potential~(\ref{PT_potential}) is equal to the minimum of the potential~(\ref{ion_Potential});

- the energy of the ground state $E_0$ is equal in both potentials. 

The found approximating potential is shown in Fig.~\ref{Potentials}b. Using Eq.~(\ref{levels_energy}) we find the energy of the first excited state $E_1=-18.6$~eV. 

\subsection{1D helium. Calculation of the electronic density spatial distribution in the AIS.}
\label{spatial}
Setting $b \approx a$ in potential~(\ref{eq_Potential}) we consider a two-electron atom. The parameter $b$ is chosen so that the atom ionization energy is close to the one of real helium. The found value is $b=1/\sqrt{3}=0.577$. The energy of the atomic ground state is the energy of the ionic ground state (-54.4 eV) minus the ionization energy 24.9eV (24.6eV in actual helium); the found energy of the ground state is -79.3 eV. 

\begin{figure}[]
\centering
\includegraphics[width=0.9\linewidth]{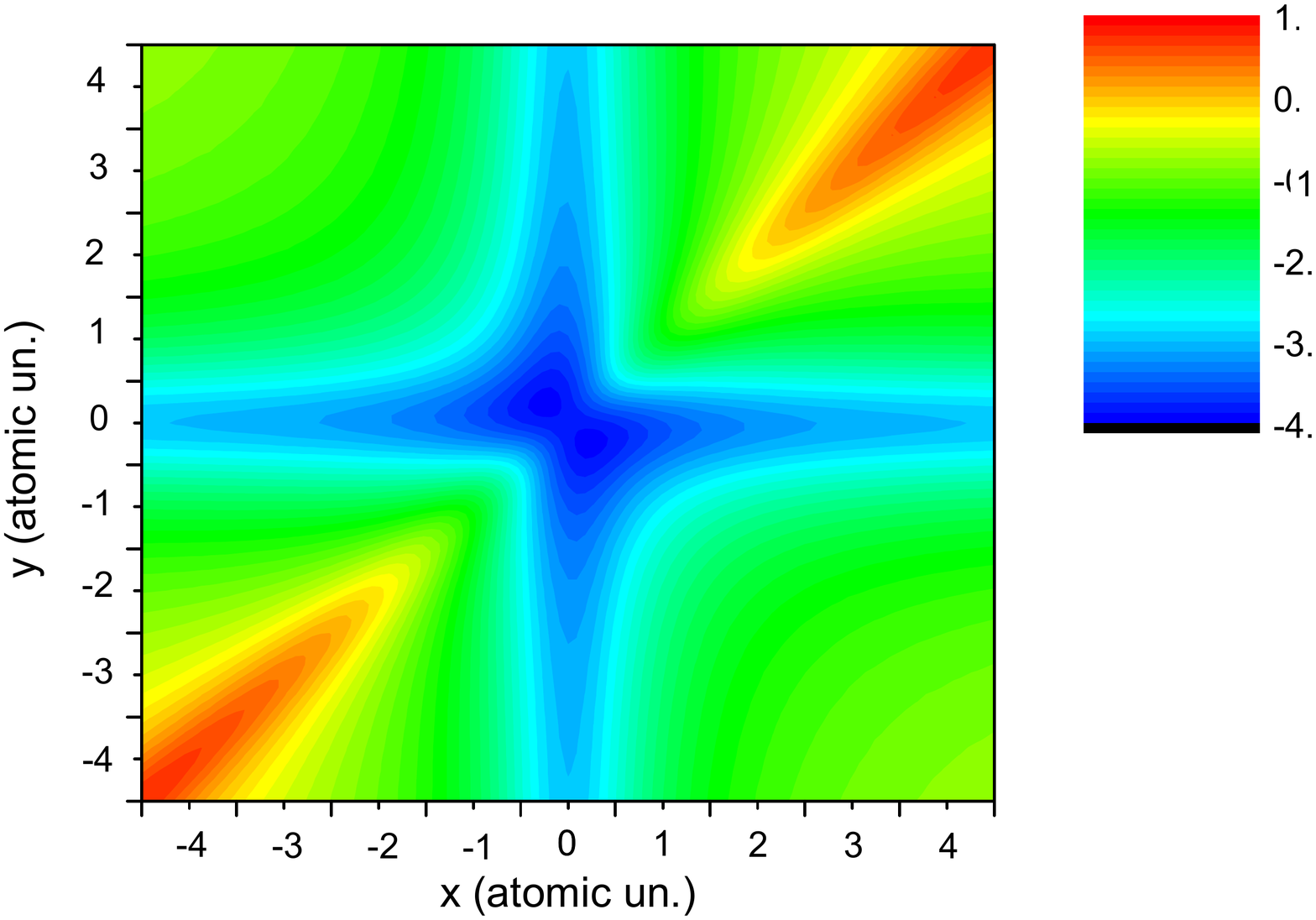}
\caption{Potential given by Eq.~(\ref{eq_Potential}) describing the model 1D helium atom.}
\label{Potential3D}
\end{figure}

The potential given by Eq.~(\ref{eq_Potential}) under $a=1/\sqrt{2}$ and $b=1/\sqrt{3}$ is shown in Fig.~\ref{Potential3D}. 

To find analytically the energy of (1,1) state in zero approximation one can neglect the electronic interaction. Within this approximation the energy of the state is doubled energy $E_1$ given by eq.~(\ref{levels_energy}); this gives $2 \times -18.6$eV$=-37.2$eV. This is pretty close to the level energy of $-34.0$ eV found numerically solving TDSE. This similarity can be explained as follows: for $n \ge 1$ the 1D soft-Coulomb potential's levels are lower than the ones of the modified P\"oschl-Teller potential. However, the electronic repulsion in the atom moves them up. So finally these two "shifts" partly compensate each other for $n=1$.

The wave-functions for several states found via numerical TDSE solution are shown in Fig.~\ref{Wave-functions}. 

The wave-functions and the energies of the bound states (the two upper rows in the figure) are found similarly to the case of the hydrogen-like ion: the initial wave-function for a state $(n_x, n_y)$ is set as a symmetrized product of the one-electron wave-functions $(n_x)$ and $(n_x)$ in the modified P\"oschl-Teller potential; TDSE is solved for a long enough time, so that the eigenstate survives and the remaining part of the initial wave-function is absorbed after spreading. The maximum in the energy distribution~(\ref{energy_distribution}) gives the eigenstate energy. 

However, this procedure could not be applied directly for an AIS: this state decays, so, after some time we find zero population in all states. So the procedure is slightly modified: first we set the initial wave-function (1,1) as described above, then propagate the TDSE and find the energy $E_{AIS}$ of the AIS; then, using the wave-function found during the TDSE integration $\psi(x,y,t)$ we calculate spectrum $\psi(x,y,\omega)$ and finally find the spectral component with this energy: $\Psi(x,y) \equiv \psi(x,y, \omega=E_{AIS})$. This function is shown in Fig.~\ref{Wave-functions}, bottom row. 
\typeout{}
\bibliography{lit.bib}

\begin{thebibliography}{50}%
\makeatletter
\providecommand \@ifxundefined [1]{%
 \@ifx{#1\undefined}
}%
\providecommand \@ifnum [1]{%
 \ifnum #1\expandafter \@firstoftwo
 \else \expandafter \@secondoftwo
 \fi
}%
\providecommand \@ifx [1]{%
 \ifx #1\expandafter \@firstoftwo
 \else \expandafter \@secondoftwo
 \fi
}%
\providecommand \natexlab [1]{#1}%
\providecommand \enquote  [1]{``#1''}%
\providecommand \bibnamefont  [1]{#1}%
\providecommand \bibfnamefont [1]{#1}%
\providecommand \citenamefont [1]{#1}%
\providecommand \href@noop [0]{\@secondoftwo}%
\providecommand \href [0]{\begingroup \@sanitize@url \@href}%
\providecommand \@href[1]{\@@startlink{#1}\@@href}%
\providecommand \@@href[1]{\endgroup#1\@@endlink}%
\providecommand \@sanitize@url [0]{\catcode `\\12\catcode `\$12\catcode
  `\&12\catcode `\#12\catcode `\^12\catcode `\_12\catcode `\%12\relax}%
\providecommand \@@startlink[1]{}%
\providecommand \@@endlink[0]{}%
\providecommand \url  [0]{\begingroup\@sanitize@url \@url }%
\providecommand \@url [1]{\endgroup\@href {#1}{\urlprefix }}%
\providecommand \urlprefix  [0]{URL }%
\providecommand \Eprint [0]{\href }%
\providecommand \doibase [0]{https://doi.org/}%
\providecommand \selectlanguage [0]{\@gobble}%
\providecommand \bibinfo  [0]{\@secondoftwo}%
\providecommand \bibfield  [0]{\@secondoftwo}%
\providecommand \translation [1]{[#1]}%
\providecommand \BibitemOpen [0]{}%
\providecommand \bibitemStop [0]{}%
\providecommand \bibitemNoStop [0]{.\EOS\space}%
\providecommand \EOS [0]{\spacefactor3000\relax}%
\providecommand \BibitemShut  [1]{\csname bibitem#1\endcsname}%
\let\auto@bib@innerbib\@empty
\bibitem [{\citenamefont {Bethe}\ and\ \citenamefont
  {Salpeter}(1957)}]{Bete_Salpeter}%
  \BibitemOpen
  \bibfield  {author} {\bibinfo {author} {\bibfnamefont {H.}~\bibnamefont
  {Bethe}}\ and\ \bibinfo {author} {\bibfnamefont {E.}~\bibnamefont
  {Salpeter}},\ }\href {https://doi.org/10.1007/978-1-4613-4104-8} {\emph
  {\bibinfo {title} {Quantum Mechanics of One-and-Two-Electron Atoms}}},\
  \bibinfo {edition} {1st}\ ed.\ (\bibinfo  {publisher} {Springer New York,
  NY},\ \bibinfo {year} {1957})\BibitemShut {NoStop}%
\bibitem [{\citenamefont {Tanner}\ \emph {et~al.}(2000)\citenamefont {Tanner},
  \citenamefont {Richter},\ and\ \citenamefont
  {Rost}}]{Review_two-electron_atoms}%
  \BibitemOpen
  \bibfield  {author} {\bibinfo {author} {\bibfnamefont {G.}~\bibnamefont
  {Tanner}}, \bibinfo {author} {\bibfnamefont {K.}~\bibnamefont {Richter}},\
  and\ \bibinfo {author} {\bibfnamefont {J.-M.}\ \bibnamefont {Rost}},\
  }\bibfield  {title} {\bibinfo {title} {The theory of two-electron atoms:
  between ground state and complete fragmentation},\ }\href
  {https://doi.org/10.1103/RevModPhys.72.497} {\bibfield  {journal} {\bibinfo
  {journal} {Rev. Mod. Phys.}\ }\textbf {\bibinfo {volume} {72}},\ \bibinfo
  {pages} {497} (\bibinfo {year} {2000})}\BibitemShut {NoStop}%
\bibitem [{\citenamefont {Smyth}\ \emph {et~al.}(1998)\citenamefont {Smyth},
  \citenamefont {Parker},\ and\ \citenamefont {Taylor}}]{Smyth1998}%
  \BibitemOpen
  \bibfield  {author} {\bibinfo {author} {\bibfnamefont {E.~S.}\ \bibnamefont
  {Smyth}}, \bibinfo {author} {\bibfnamefont {J.~S.}\ \bibnamefont {Parker}},\
  and\ \bibinfo {author} {\bibfnamefont {K.}~\bibnamefont {Taylor}},\
  }\bibfield  {title} {\bibinfo {title} {Numerical integration of the
  time-dependent schrödinger equation for laser-driven helium},\ }\href
  {https://doi.org/https://doi.org/10.1016/S0010-4655(98)00083-6} {\bibfield
  {journal} {\bibinfo  {journal} {Computer Physics Communications}\ }\textbf
  {\bibinfo {volume} {114}},\ \bibinfo {pages} {1} (\bibinfo {year}
  {1998})}\BibitemShut {NoStop}%
\bibitem [{\citenamefont {Ruiz}\ \emph {et~al.}(2006)\citenamefont {Ruiz},
  \citenamefont {Plaja}, \citenamefont {Roso},\ and\ \citenamefont
  {Becker}}]{Ruiz2006}%
  \BibitemOpen
  \bibfield  {author} {\bibinfo {author} {\bibfnamefont {C.}~\bibnamefont
  {Ruiz}}, \bibinfo {author} {\bibfnamefont {L.}~\bibnamefont {Plaja}},
  \bibinfo {author} {\bibfnamefont {L.}~\bibnamefont {Roso}},\ and\ \bibinfo
  {author} {\bibfnamefont {A.}~\bibnamefont {Becker}},\ }\bibfield  {title}
  {\bibinfo {title} {Ab initio calculation of the double ionization of helium
  in a few-cycle laser pulse beyond the one-dimensional approximation},\ }\href
  {https://doi.org/10.1103/PhysRevLett.96.053001} {\bibfield  {journal}
  {\bibinfo  {journal} {Phys. Rev. Lett.}\ }\textbf {\bibinfo {volume} {96}},\
  \bibinfo {pages} {053001} (\bibinfo {year} {2006})}\BibitemShut {NoStop}%
\bibitem [{\citenamefont {Nogami}\ \emph {et~al.}(1976)\citenamefont {Nogami},
  \citenamefont {Vallières},\ and\ \citenamefont {van Dijk}}]{1D_He_1976}%
  \BibitemOpen
  \bibfield  {author} {\bibinfo {author} {\bibfnamefont {Y.}~\bibnamefont
  {Nogami}}, \bibinfo {author} {\bibfnamefont {M.}~\bibnamefont {Vallières}},\
  and\ \bibinfo {author} {\bibfnamefont {W.}~\bibnamefont {van Dijk}},\
  }\bibfield  {title} {\bibinfo {title} {Hartree–fock approximation for the
  one‐dimensional ’’helium atom’’},\ }\href
  {https://doi.org/10.1119/1.10291} {\bibfield  {journal} {\bibinfo  {journal}
  {American Journal of Physics}\ }\textbf {\bibinfo {volume} {44}},\ \bibinfo
  {pages} {886} (\bibinfo {year} {1976})},\ \Eprint
  {https://arxiv.org/abs/https://doi.org/10.1119/1.10291}
  {https://doi.org/10.1119/1.10291} \BibitemShut {NoStop}%
\bibitem [{\citenamefont {Harriss}\ and\ \citenamefont
  {Rioux}(1980)}]{1d_He_1980}%
  \BibitemOpen
  \bibfield  {author} {\bibinfo {author} {\bibfnamefont {D.~K.}\ \bibnamefont
  {Harriss}}\ and\ \bibinfo {author} {\bibfnamefont {F.}~\bibnamefont
  {Rioux}},\ }\bibfield  {title} {\bibinfo {title} {A simple hartree scf
  calculation on a one-dimensional model of the he atom},\ }\href
  {https://doi.org/10.1021/ed057p491} {\bibfield  {journal} {\bibinfo
  {journal} {Journal of Chemical Education}\ }\textbf {\bibinfo {volume}
  {57}},\ \bibinfo {pages} {491} (\bibinfo {year} {1980})},\ \Eprint
  {https://arxiv.org/abs/https://doi.org/10.1021/ed057p491}
  {https://doi.org/10.1021/ed057p491} \BibitemShut {NoStop}%
\bibitem [{\citenamefont {L{\'{o}}pez-Castillo}\ \emph
  {et~al.}(1996)\citenamefont {L{\'{o}}pez-Castillo}, \citenamefont
  {de~Aguiar},\ and\ \citenamefont {de~Almeida}}]{1D_He_1996}%
  \BibitemOpen
  \bibfield  {author} {\bibinfo {author} {\bibfnamefont {A.}~\bibnamefont
  {L{\'{o}}pez-Castillo}}, \bibinfo {author} {\bibfnamefont {M.~A.~M.}\
  \bibnamefont {de~Aguiar}},\ and\ \bibinfo {author} {\bibfnamefont {A.~M.~O.}\
  \bibnamefont {de~Almeida}},\ }\bibfield  {title} {\bibinfo {title} {On the
  one-dimensional helium atom},\ }\href
  {https://doi.org/10.1088/0953-4075/29/2/009} {\bibfield  {journal} {\bibinfo
  {journal} {Journal of Physics B: Atomic, Molecular and Optical Physics}\
  }\textbf {\bibinfo {volume} {29}},\ \bibinfo {pages} {197} (\bibinfo {year}
  {1996})}\BibitemShut {NoStop}%
\bibitem [{\citenamefont {Skobelev}(2018)}]{1D_He_2018}%
  \BibitemOpen
  \bibfield  {author} {\bibinfo {author} {\bibfnamefont {V.}~\bibnamefont
  {Skobelev}},\ }\bibfield  {title} {\bibinfo {title} {Ground state energy of a
  one-dimensional helium atom.},\ }\href
  {https://doi.org/10.1007/s11182-018-1473-8} {\bibfield  {journal} {\bibinfo
  {journal} {Russ Phys J}\ }\textbf {\bibinfo {volume} {61}},\ \bibinfo {pages}
  {887} (\bibinfo {year} {2018})},\ \Eprint
  {https://arxiv.org/abs/https://doi.org/10.1007/s11182-018-1473-8}
  {https://doi.org/10.1007/s11182-018-1473-8} \BibitemShut {NoStop}%
\bibitem [{\citenamefont {Grobe}\ and\ \citenamefont
  {Eberly}(1992)}]{Grobe1992}%
  \BibitemOpen
  \bibfield  {author} {\bibinfo {author} {\bibfnamefont {R.}~\bibnamefont
  {Grobe}}\ and\ \bibinfo {author} {\bibfnamefont {J.~H.}\ \bibnamefont
  {Eberly}},\ }\bibfield  {title} {\bibinfo {title} {Photoelectron spectra for
  a two-electron system in a strong laser field},\ }\href
  {https://doi.org/10.1103/PhysRevLett.68.2905} {\bibfield  {journal} {\bibinfo
   {journal} {Phys. Rev. Lett.}\ }\textbf {\bibinfo {volume} {68}},\ \bibinfo
  {pages} {2905} (\bibinfo {year} {1992})}\BibitemShut {NoStop}%
\bibitem [{\citenamefont {Haan}\ \emph {et~al.}(1994)\citenamefont {Haan},
  \citenamefont {Grobe},\ and\ \citenamefont {Eberly}}]{Haan1994}%
  \BibitemOpen
  \bibfield  {author} {\bibinfo {author} {\bibfnamefont {S.~L.}\ \bibnamefont
  {Haan}}, \bibinfo {author} {\bibfnamefont {R.}~\bibnamefont {Grobe}},\ and\
  \bibinfo {author} {\bibfnamefont {J.~H.}\ \bibnamefont {Eberly}},\ }\bibfield
   {title} {\bibinfo {title} {Numerical study of autoionizing states in
  completely correlated two-electron systems},\ }\href
  {https://doi.org/10.1103/PhysRevA.50.378} {\bibfield  {journal} {\bibinfo
  {journal} {Phys. Rev. A}\ }\textbf {\bibinfo {volume} {50}},\ \bibinfo
  {pages} {378} (\bibinfo {year} {1994})}\BibitemShut {NoStop}%
\bibitem [{\citenamefont {Bauer}(1997)}]{Bauer1997}%
  \BibitemOpen
  \bibfield  {author} {\bibinfo {author} {\bibfnamefont {D.}~\bibnamefont
  {Bauer}},\ }\bibfield  {title} {\bibinfo {title} {Two-dimensional,
  two-electron model atom in a laser pulse: Exact treatment,
  single-active-electron analysis, time-dependent density-functional theory,
  classical calculations, and nonsequential ionization},\ }\href
  {https://doi.org/10.1103/PhysRevA.56.3028} {\bibfield  {journal} {\bibinfo
  {journal} {Phys. Rev. A}\ }\textbf {\bibinfo {volume} {56}},\ \bibinfo
  {pages} {3028} (\bibinfo {year} {1997})}\BibitemShut {NoStop}%
\bibitem [{\citenamefont {Lappas}\ and\ \citenamefont {van
  Leeuwen}(1998)}]{Lappas_1998}%
  \BibitemOpen
  \bibfield  {author} {\bibinfo {author} {\bibfnamefont {D.~G.}\ \bibnamefont
  {Lappas}}\ and\ \bibinfo {author} {\bibfnamefont {R.}~\bibnamefont {van
  Leeuwen}},\ }\bibfield  {title} {\bibinfo {title} {Electron correlation
  effects in the double ionization of he},\ }\href
  {https://doi.org/10.1088/0953-4075/31/6/001} {\bibfield  {journal} {\bibinfo
  {journal} {Journal of Physics B: Atomic, Molecular and Optical Physics}\
  }\textbf {\bibinfo {volume} {31}},\ \bibinfo {pages} {L249} (\bibinfo {year}
  {1998})}\BibitemShut {NoStop}%
\bibitem [{\citenamefont {Lein}\ \emph {et~al.}(2000)\citenamefont {Lein},
  \citenamefont {Gross},\ and\ \citenamefont {Engel}}]{Lein2000}%
  \BibitemOpen
  \bibfield  {author} {\bibinfo {author} {\bibfnamefont {M.}~\bibnamefont
  {Lein}}, \bibinfo {author} {\bibfnamefont {E.~K.~U.}\ \bibnamefont {Gross}},\
  and\ \bibinfo {author} {\bibfnamefont {V.}~\bibnamefont {Engel}},\ }\bibfield
   {title} {\bibinfo {title} {Intense-field double ionization of helium:
  Identifying the mechanism},\ }\href
  {https://doi.org/10.1103/PhysRevLett.85.4707} {\bibfield  {journal} {\bibinfo
   {journal} {Phys. Rev. Lett.}\ }\textbf {\bibinfo {volume} {85}},\ \bibinfo
  {pages} {4707} (\bibinfo {year} {2000})}\BibitemShut {NoStop}%
\bibitem [{\citenamefont {Borisova}\ \emph {et~al.}(2020)\citenamefont
  {Borisova}, \citenamefont {Stoo{\ss}}, \citenamefont {Dingeldey},
  \citenamefont {Kaldun}, \citenamefont {Ding}, \citenamefont {Birk},
  \citenamefont {Hartmann}, \citenamefont {Heldt}, \citenamefont {Ott},\ and\
  \citenamefont {Pfeifer}}]{Borisova2020}%
  \BibitemOpen
  \bibfield  {author} {\bibinfo {author} {\bibfnamefont {G.~D.}\ \bibnamefont
  {Borisova}}, \bibinfo {author} {\bibfnamefont {V.}~\bibnamefont {Stoo{\ss}}},
  \bibinfo {author} {\bibfnamefont {A.}~\bibnamefont {Dingeldey}}, \bibinfo
  {author} {\bibfnamefont {A.}~\bibnamefont {Kaldun}}, \bibinfo {author}
  {\bibfnamefont {T.}~\bibnamefont {Ding}}, \bibinfo {author} {\bibfnamefont
  {P.}~\bibnamefont {Birk}}, \bibinfo {author} {\bibfnamefont {M.}~\bibnamefont
  {Hartmann}}, \bibinfo {author} {\bibfnamefont {T.}~\bibnamefont {Heldt}},
  \bibinfo {author} {\bibfnamefont {C.}~\bibnamefont {Ott}},\ and\ \bibinfo
  {author} {\bibfnamefont {T.}~\bibnamefont {Pfeifer}},\ }\bibfield  {title}
  {\bibinfo {title} {Strong-field-induced single and double ionization dynamics
  from single and double excitations in a two-electron atom},\ }\href
  {https://doi.org/10.1088/2399-6528/ab6175} {\bibfield  {journal} {\bibinfo
  {journal} {Journal of Physics Communications}\ }\textbf {\bibinfo {volume}
  {4}},\ \bibinfo {pages} {055012} (\bibinfo {year} {2020})}\BibitemShut
  {NoStop}%
\bibitem [{\citenamefont {Volkova}\ \emph {et~al.}(2004)\citenamefont
  {Volkova}, \citenamefont {Gridchin}, \citenamefont {Popov},\ and\
  \citenamefont {Tikhonova}}]{Volkova2004}%
  \BibitemOpen
  \bibfield  {author} {\bibinfo {author} {\bibfnamefont {E.~A.}\ \bibnamefont
  {Volkova}}, \bibinfo {author} {\bibfnamefont {V.~V.}\ \bibnamefont
  {Gridchin}}, \bibinfo {author} {\bibfnamefont {A.~M.}\ \bibnamefont
  {Popov}},\ and\ \bibinfo {author} {\bibfnamefont {O.~V.}\ \bibnamefont
  {Tikhonova}},\ }\bibfield  {title} {\bibinfo {title} {Ionization and
  stabilization of a two-electron atom in a strong electromagnetic field},\
  }\href {https://doi.org/10.1134/1.1800188} {\bibfield  {journal} {\bibinfo
  {journal} {J Exp Theor Phys}\ }\textbf {\bibinfo {volume} {99}},\ \bibinfo
  {pages} {320} (\bibinfo {year} {2004})}\BibitemShut {NoStop}%
\bibitem [{\citenamefont {Roso}\ \emph {et~al.}(2005)\citenamefont {Roso},
  \citenamefont {Plaja}, \citenamefont {Moreno}, \citenamefont {Jarque},
  \citenamefont {Vazquez~de Aldana}, \citenamefont {San~Román},\ and\
  \citenamefont {Ruiz}}]{Roso2005}%
  \BibitemOpen
  \bibfield  {author} {\bibinfo {author} {\bibfnamefont {L.}~\bibnamefont
  {Roso}}, \bibinfo {author} {\bibfnamefont {L.}~\bibnamefont {Plaja}},
  \bibinfo {author} {\bibfnamefont {P.}~\bibnamefont {Moreno}}, \bibinfo
  {author} {\bibfnamefont {E.}~\bibnamefont {Jarque}}, \bibinfo {author}
  {\bibfnamefont {J.}~\bibnamefont {Vazquez~de Aldana}}, \bibinfo {author}
  {\bibfnamefont {J.}~\bibnamefont {San~Román}},\ and\ \bibinfo {author}
  {\bibfnamefont {C.}~\bibnamefont {Ruiz}},\ }\bibfield  {title} {\bibinfo
  {title} {Multielectron atomic models using the rochester one-dimensional
  potential},\ }\href@noop {} {\bibfield  {journal} {\bibinfo  {journal} {Laser
  Physics}\ }\textbf {\bibinfo {volume} {15}},\ \bibinfo {pages} {1393–1409}
  (\bibinfo {year} {2005})}\BibitemShut {NoStop}%
\bibitem [{\citenamefont {Zhao}\ and\ \citenamefont {Lein}(2012)}]{Zhao2012}%
  \BibitemOpen
  \bibfield  {author} {\bibinfo {author} {\bibfnamefont {J.}~\bibnamefont
  {Zhao}}\ and\ \bibinfo {author} {\bibfnamefont {M.}~\bibnamefont {Lein}},\
  }\bibfield  {title} {\bibinfo {title} {Probing fano resonances with
  ultrashort pulses},\ }\href {https://doi.org/10.1088/1367-2630/14/6/065003}
  {\bibfield  {journal} {\bibinfo  {journal} {New Journal of Physics}\ }\textbf
  {\bibinfo {volume} {14}},\ \bibinfo {pages} {065003} (\bibinfo {year}
  {2012})}\BibitemShut {NoStop}%
\bibitem [{\citenamefont {Koval}\ \emph {et~al.}(2007)\citenamefont {Koval},
  \citenamefont {Wilken}, \citenamefont {Bauer},\ and\ \citenamefont
  {Keitel}}]{Koval2007}%
  \BibitemOpen
  \bibfield  {author} {\bibinfo {author} {\bibfnamefont {P.}~\bibnamefont
  {Koval}}, \bibinfo {author} {\bibfnamefont {F.}~\bibnamefont {Wilken}},
  \bibinfo {author} {\bibfnamefont {D.}~\bibnamefont {Bauer}},\ and\ \bibinfo
  {author} {\bibfnamefont {C.~H.}\ \bibnamefont {Keitel}},\ }\bibfield  {title}
  {\bibinfo {title} {Nonsequential double recombination in intense laser
  fields},\ }\href {https://doi.org/10.1103/PhysRevLett.98.043904} {\bibfield
  {journal} {\bibinfo  {journal} {Phys. Rev. Lett.}\ }\textbf {\bibinfo
  {volume} {98}},\ \bibinfo {pages} {043904} (\bibinfo {year}
  {2007})}\BibitemShut {NoStop}%
\bibitem [{\citenamefont {Efimov}\ \emph {et~al.}(2018)\citenamefont {Efimov},
  \citenamefont {Maksymov}, \citenamefont {Prauzner-Bechcicki}, \citenamefont
  {Thiede}, \citenamefont {Eckhardt}, \citenamefont {Chac\'on}, \citenamefont
  {Lewenstein},\ and\ \citenamefont {Zakrzewski}}]{Efimov2018}%
  \BibitemOpen
  \bibfield  {author} {\bibinfo {author} {\bibfnamefont {D.~K.}\ \bibnamefont
  {Efimov}}, \bibinfo {author} {\bibfnamefont {A.}~\bibnamefont {Maksymov}},
  \bibinfo {author} {\bibfnamefont {J.~S.}\ \bibnamefont {Prauzner-Bechcicki}},
  \bibinfo {author} {\bibfnamefont {J.~H.}\ \bibnamefont {Thiede}}, \bibinfo
  {author} {\bibfnamefont {B.}~\bibnamefont {Eckhardt}}, \bibinfo {author}
  {\bibfnamefont {A.}~\bibnamefont {Chac\'on}}, \bibinfo {author}
  {\bibfnamefont {M.}~\bibnamefont {Lewenstein}},\ and\ \bibinfo {author}
  {\bibfnamefont {J.}~\bibnamefont {Zakrzewski}},\ }\bibfield  {title}
  {\bibinfo {title} {Restricted-space ab initio models for double ionization by
  strong laser pulses},\ }\href {https://doi.org/10.1103/PhysRevA.98.013405}
  {\bibfield  {journal} {\bibinfo  {journal} {Phys. Rev. A}\ }\textbf {\bibinfo
  {volume} {98}},\ \bibinfo {pages} {013405} (\bibinfo {year}
  {2018})}\BibitemShut {NoStop}%
\bibitem [{\citenamefont {Milošević}(2007)}]{Milosevic_2007}%
  \BibitemOpen
  \bibfield  {author} {\bibinfo {author} {\bibfnamefont {D.~B.}\ \bibnamefont
  {Milošević}},\ }\bibfield  {title} {\bibinfo {title} {High-energy
  stimulated emission from plasma ablation pumped by resonant high-order
  harmonic generation},\ }\href {https://doi.org/10.1088/0953-4075/40/17/005}
  {\bibfield  {journal} {\bibinfo  {journal} {Journal of Physics B: Atomic,
  Molecular and Optical Physics}\ }\textbf {\bibinfo {volume} {40}},\ \bibinfo
  {pages} {3367} (\bibinfo {year} {2007})}\BibitemShut {NoStop}%
\bibitem [{\citenamefont {Ivanov}\ and\ \citenamefont
  {Kheifets}(2008)}]{Kheifets2008}%
  \BibitemOpen
  \bibfield  {author} {\bibinfo {author} {\bibfnamefont {I.~A.}\ \bibnamefont
  {Ivanov}}\ and\ \bibinfo {author} {\bibfnamefont {A.~S.}\ \bibnamefont
  {Kheifets}},\ }\bibfield  {title} {\bibinfo {title} {Resonant enhancement of
  generation of harmonics},\ }\href
  {https://doi.org/10.1103/PhysRevA.78.053406} {\bibfield  {journal} {\bibinfo
  {journal} {Phys. Rev. A}\ }\textbf {\bibinfo {volume} {78}},\ \bibinfo
  {pages} {053406} (\bibinfo {year} {2008})}\BibitemShut {NoStop}%
\bibitem [{\citenamefont {Strelkov}(2010)}]{Strelkov2010}%
  \BibitemOpen
  \bibfield  {author} {\bibinfo {author} {\bibfnamefont {V.}~\bibnamefont
  {Strelkov}},\ }\bibfield  {title} {\bibinfo {title} {Role of autoionizing
  state in resonant high-order harmonic generation and attosecond pulse
  production},\ }\href {https://doi.org/10.1103/PhysRevLett.104.123901}
  {\bibfield  {journal} {\bibinfo  {journal} {Phys. Rev. Lett.}\ }\textbf
  {\bibinfo {volume} {104}},\ \bibinfo {pages} {123901} (\bibinfo {year}
  {2010})}\BibitemShut {NoStop}%
\bibitem [{\citenamefont {Frolov}\ \emph {et~al.}(2010)\citenamefont {Frolov},
  \citenamefont {Manakov},\ and\ \citenamefont {Starace}}]{Frolov2010}%
  \BibitemOpen
  \bibfield  {author} {\bibinfo {author} {\bibfnamefont {M.~V.}\ \bibnamefont
  {Frolov}}, \bibinfo {author} {\bibfnamefont {N.~L.}\ \bibnamefont
  {Manakov}},\ and\ \bibinfo {author} {\bibfnamefont {A.~F.}\ \bibnamefont
  {Starace}},\ }\bibfield  {title} {\bibinfo {title} {Potential barrier effects
  in high-order harmonic generation by transition-metal ions},\ }\href
  {https://doi.org/10.1103/PhysRevA.82.023424} {\bibfield  {journal} {\bibinfo
  {journal} {Phys. Rev. A}\ }\textbf {\bibinfo {volume} {82}},\ \bibinfo
  {pages} {023424} (\bibinfo {year} {2010})}\BibitemShut {NoStop}%
\bibitem [{\citenamefont {Strelkov}\ \emph {et~al.}(2014)\citenamefont
  {Strelkov}, \citenamefont {Khokhlova},\ and\ \citenamefont
  {Shubin}}]{Strelkov2014}%
  \BibitemOpen
  \bibfield  {author} {\bibinfo {author} {\bibfnamefont {V.~V.}\ \bibnamefont
  {Strelkov}}, \bibinfo {author} {\bibfnamefont {M.~A.}\ \bibnamefont
  {Khokhlova}},\ and\ \bibinfo {author} {\bibfnamefont {N.~Y.}\ \bibnamefont
  {Shubin}},\ }\bibfield  {title} {\bibinfo {title} {High-order harmonic
  generation and fano resonances},\ }\href
  {https://doi.org/10.1103/PhysRevA.89.053833} {\bibfield  {journal} {\bibinfo
  {journal} {Phys. Rev. A}\ }\textbf {\bibinfo {volume} {89}},\ \bibinfo
  {pages} {053833} (\bibinfo {year} {2014})}\BibitemShut {NoStop}%
\bibitem [{\citenamefont {Wahyutama}\ \emph {et~al.}(2019)\citenamefont
  {Wahyutama}, \citenamefont {Sato},\ and\ \citenamefont
  {Ishikawa}}]{Wahyutama2019}%
  \BibitemOpen
  \bibfield  {author} {\bibinfo {author} {\bibfnamefont {I.~S.}\ \bibnamefont
  {Wahyutama}}, \bibinfo {author} {\bibfnamefont {T.}~\bibnamefont {Sato}},\
  and\ \bibinfo {author} {\bibfnamefont {K.~L.}\ \bibnamefont {Ishikawa}},\
  }\bibfield  {title} {\bibinfo {title} {Time-dependent multiconfiguration
  self-consistent-field study on resonantly enhanced high-order harmonic
  generation from transition-metal elements},\ }\href
  {https://doi.org/10.1103/PhysRevA.99.063420} {\bibfield  {journal} {\bibinfo
  {journal} {Phys. Rev. A}\ }\textbf {\bibinfo {volume} {99}},\ \bibinfo
  {pages} {063420} (\bibinfo {year} {2019})}\BibitemShut {NoStop}%
\bibitem [{\citenamefont {Ganeev}\ \emph {et~al.}(2006)\citenamefont {Ganeev},
  \citenamefont {Suzuki}, \citenamefont {Baba}, \citenamefont {Kuroda},\ and\
  \citenamefont {Ozaki}}]{Ganeev2006}%
  \BibitemOpen
  \bibfield  {author} {\bibinfo {author} {\bibfnamefont {R.~A.}\ \bibnamefont
  {Ganeev}}, \bibinfo {author} {\bibfnamefont {M.}~\bibnamefont {Suzuki}},
  \bibinfo {author} {\bibfnamefont {M.}~\bibnamefont {Baba}}, \bibinfo {author}
  {\bibfnamefont {H.}~\bibnamefont {Kuroda}},\ and\ \bibinfo {author}
  {\bibfnamefont {T.}~\bibnamefont {Ozaki}},\ }\bibfield  {title} {\bibinfo
  {title} {Strong resonance enhancement of a single harmonic generated in the
  extreme ultraviolet range},\ }\href {https://doi.org/10.1364/OL.31.001699}
  {\bibfield  {journal} {\bibinfo  {journal} {Opt. Lett.}\ }\textbf {\bibinfo
  {volume} {31}},\ \bibinfo {pages} {1699} (\bibinfo {year}
  {2006})}\BibitemShut {NoStop}%
\bibitem [{\citenamefont {Gilbertson}\ \emph {et~al.}(2008)\citenamefont
  {Gilbertson}, \citenamefont {Mashiko}, \citenamefont {Li}, \citenamefont
  {Moon},\ and\ \citenamefont {Chang}}]{Chang2008}%
  \BibitemOpen
  \bibfield  {author} {\bibinfo {author} {\bibfnamefont {S.}~\bibnamefont
  {Gilbertson}}, \bibinfo {author} {\bibfnamefont {H.}~\bibnamefont {Mashiko}},
  \bibinfo {author} {\bibfnamefont {C.}~\bibnamefont {Li}}, \bibinfo {author}
  {\bibfnamefont {E.}~\bibnamefont {Moon}},\ and\ \bibinfo {author}
  {\bibfnamefont {Z.}~\bibnamefont {Chang}},\ }\bibfield  {title} {\bibinfo
  {title} {Effects of laser pulse duration on extreme ultraviolet spectra from
  double optical gating},\ }\href {https://doi.org/10.1063/1.2982589}
  {\bibfield  {journal} {\bibinfo  {journal} {Applied Physics Letters}\
  }\textbf {\bibinfo {volume} {93}},\ \bibinfo {pages} {111105} (\bibinfo
  {year} {2008})},\ \Eprint
  {https://arxiv.org/abs/https://doi.org/10.1063/1.2982589}
  {https://doi.org/10.1063/1.2982589} \BibitemShut {NoStop}%
\bibitem [{\citenamefont {Singh}\ \emph {et~al.}(2021)\citenamefont {Singh},
  \citenamefont {Fareed}, \citenamefont {Strelkov}, \citenamefont
  {Grum-Grzhimailo}, \citenamefont {Magunov}, \citenamefont {Laram\'{e}e},
  \citenamefont {L\'{e}gar\'{e}},\ and\ \citenamefont {Ozaki}}]{Singh2021}%
  \BibitemOpen
  \bibfield  {author} {\bibinfo {author} {\bibfnamefont {M.}~\bibnamefont
  {Singh}}, \bibinfo {author} {\bibfnamefont {M.~A.}\ \bibnamefont {Fareed}},
  \bibinfo {author} {\bibfnamefont {V.}~\bibnamefont {Strelkov}}, \bibinfo
  {author} {\bibfnamefont {A.~N.}\ \bibnamefont {Grum-Grzhimailo}}, \bibinfo
  {author} {\bibfnamefont {A.}~\bibnamefont {Magunov}}, \bibinfo {author}
  {\bibfnamefont {A.}~\bibnamefont {Laram\'{e}e}}, \bibinfo {author}
  {\bibfnamefont {F.}~\bibnamefont {L\'{e}gar\'{e}}},\ and\ \bibinfo {author}
  {\bibfnamefont {T.}~\bibnamefont {Ozaki}},\ }\bibfield  {title} {\bibinfo
  {title} {Intense quasi-monochromatic resonant harmonic generation in the
  multiphoton ionization regime},\ }\href
  {https://doi.org/10.1364/OPTICA.434185} {\bibfield  {journal} {\bibinfo
  {journal} {Optica}\ }\textbf {\bibinfo {volume} {8}},\ \bibinfo {pages}
  {1122} (\bibinfo {year} {2021})}\BibitemShut {NoStop}%
\bibitem [{\citenamefont {Shiner}\ \emph {et~al.}(2011)\citenamefont {Shiner},
  \citenamefont {Schmidt}, \citenamefont {Trallero-Herrero},\ and\
  \citenamefont {et~al.}}]{Shiner2011}%
  \BibitemOpen
  \bibfield  {author} {\bibinfo {author} {\bibfnamefont {A.}~\bibnamefont
  {Shiner}}, \bibinfo {author} {\bibfnamefont {B.}~\bibnamefont {Schmidt}},
  \bibinfo {author} {\bibfnamefont {C.}~\bibnamefont {Trallero-Herrero}},\ and\
  \bibinfo {author} {\bibnamefont {et~al.}},\ }\bibfield  {title} {\bibinfo
  {title} {Probing collective multi-electron dynamics in xenon with
  high-harmonic spectroscopy},\ }\href {https://doi.org/10.1038/nphys1940}
  {\bibfield  {journal} {\bibinfo  {journal} {Nature Phys}\ }\textbf {\bibinfo
  {volume} {7}},\ \bibinfo {pages} {464} (\bibinfo {year} {2011})}\BibitemShut
  {NoStop}%
\bibitem [{\citenamefont {Fareed}\ \emph {et~al.}(2018)\citenamefont {Fareed},
  \citenamefont {Strelkov}, \citenamefont {Singh}, \citenamefont {Thir\'e},
  \citenamefont {Mondal}, \citenamefont {Schmidt}, \citenamefont {L\'egar\'e},\
  and\ \citenamefont {Ozaki}}]{Fareed2018}%
  \BibitemOpen
  \bibfield  {author} {\bibinfo {author} {\bibfnamefont {M.~A.}\ \bibnamefont
  {Fareed}}, \bibinfo {author} {\bibfnamefont {V.~V.}\ \bibnamefont
  {Strelkov}}, \bibinfo {author} {\bibfnamefont {M.}~\bibnamefont {Singh}},
  \bibinfo {author} {\bibfnamefont {N.}~\bibnamefont {Thir\'e}}, \bibinfo
  {author} {\bibfnamefont {S.}~\bibnamefont {Mondal}}, \bibinfo {author}
  {\bibfnamefont {B.~E.}\ \bibnamefont {Schmidt}}, \bibinfo {author}
  {\bibfnamefont {F.}~\bibnamefont {L\'egar\'e}},\ and\ \bibinfo {author}
  {\bibfnamefont {T.}~\bibnamefont {Ozaki}},\ }\bibfield  {title} {\bibinfo
  {title} {Harmonic generation from neutral manganese atoms in the vicinity of
  the giant autoionization resonance},\ }\href
  {https://doi.org/10.1103/PhysRevLett.121.023201} {\bibfield  {journal}
  {\bibinfo  {journal} {Phys. Rev. Lett.}\ }\textbf {\bibinfo {volume} {121}},\
  \bibinfo {pages} {023201} (\bibinfo {year} {2018})}\BibitemShut {NoStop}%
\bibitem [{\citenamefont {Ganeev}(2012)}]{Ganeev2012_Review}%
  \BibitemOpen
  \bibfield  {author} {\bibinfo {author} {\bibfnamefont {R.}~\bibnamefont
  {Ganeev}},\ }\bibfield  {title} {\bibinfo {title} {Harmonic generation in
  laser-produced plasmas containing atoms, ions and clusters: a review},\
  }\href {https://doi.org/10.1080/09500340.2011.636155} {\bibfield  {journal}
  {\bibinfo  {journal} {Journal of Modern Optics}\ }\textbf {\bibinfo {volume}
  {59}},\ \bibinfo {pages} {409} (\bibinfo {year} {2012})},\ \Eprint
  {https://arxiv.org/abs/https://doi.org/10.1080/09500340.2011.636155}
  {https://doi.org/10.1080/09500340.2011.636155} \BibitemShut {NoStop}%
\bibitem [{\citenamefont {Fareed}\ \emph {et~al.}(2017)\citenamefont {Fareed},
  \citenamefont {Strelkov}, \citenamefont {Thir\'e}, \citenamefont {Mondal},
  \citenamefont {Schmidt}, \citenamefont {L\'egar\'e},\ and\ \citenamefont
  {Ozaki}}]{Fareed2017}%
  \BibitemOpen
  \bibfield  {author} {\bibinfo {author} {\bibfnamefont {M.~A.}\ \bibnamefont
  {Fareed}}, \bibinfo {author} {\bibfnamefont {V.~V.}\ \bibnamefont
  {Strelkov}}, \bibinfo {author} {\bibfnamefont {N.}~\bibnamefont {Thir\'e}},
  \bibinfo {author} {\bibfnamefont {S.}~\bibnamefont {Mondal}}, \bibinfo
  {author} {\bibfnamefont {B.~E.}\ \bibnamefont {Schmidt}}, \bibinfo {author}
  {\bibfnamefont {F.}~\bibnamefont {L\'egar\'e}},\ and\ \bibinfo {author}
  {\bibfnamefont {T.}~\bibnamefont {Ozaki}},\ }\bibfield  {title} {\bibinfo
  {title} {High-order harmonic generation from the dressed autoionizing
  states},\ }\href {https://doi.org/10.1038/ncomms16061} {\bibfield  {journal}
  {\bibinfo  {journal} {Nat Commun}\ }\textbf {\bibinfo {volume} {8}},\
  \bibinfo {pages} {16061} (\bibinfo {year} {2017})}\BibitemShut {NoStop}%
\bibitem [{\citenamefont {Magunov}\ \emph {et~al.}(2001)\citenamefont
  {Magunov}, \citenamefont {Rotter},\ and\ \citenamefont
  {Strakhova}}]{Magunov2001}%
  \BibitemOpen
  \bibfield  {author} {\bibinfo {author} {\bibfnamefont {A.~I.}\ \bibnamefont
  {Magunov}}, \bibinfo {author} {\bibfnamefont {I.}~\bibnamefont {Rotter}},\
  and\ \bibinfo {author} {\bibfnamefont {S.~I.}\ \bibnamefont {Strakhova}},\
  }\bibfield  {title} {\bibinfo {title} {Laser-induced continuum structures and
  double poles of the s-matrix},\ }\href
  {https://doi.org/10.1088/0953-4075/34/1/303} {\bibfield  {journal} {\bibinfo
  {journal} {Journal of Physics B: Atomic, Molecular and Optical Physics}\
  }\textbf {\bibinfo {volume} {34}},\ \bibinfo {pages} {29} (\bibinfo {year}
  {2001})}\BibitemShut {NoStop}%
\bibitem [{\citenamefont {Argenti}\ \emph {et~al.}(2015)\citenamefont
  {Argenti}, \citenamefont {Jim\'enez-Gal\'an}, \citenamefont {Marante},
  \citenamefont {Ott}, \citenamefont {Pfeifer},\ and\ \citenamefont
  {Mart\'{\i}n}}]{Argenti2015}%
  \BibitemOpen
  \bibfield  {author} {\bibinfo {author} {\bibfnamefont {L.}~\bibnamefont
  {Argenti}}, \bibinfo {author} {\bibfnamefont {A.}~\bibnamefont
  {Jim\'enez-Gal\'an}}, \bibinfo {author} {\bibfnamefont {C.}~\bibnamefont
  {Marante}}, \bibinfo {author} {\bibfnamefont {C.}~\bibnamefont {Ott}},
  \bibinfo {author} {\bibfnamefont {T.}~\bibnamefont {Pfeifer}},\ and\ \bibinfo
  {author} {\bibfnamefont {F.}~\bibnamefont {Mart\'{\i}n}},\ }\bibfield
  {title} {\bibinfo {title} {Dressing effects in the attosecond transient
  absorption spectra of doubly excited states in helium},\ }\href
  {https://doi.org/10.1103/PhysRevA.91.061403} {\bibfield  {journal} {\bibinfo
  {journal} {Phys. Rev. A}\ }\textbf {\bibinfo {volume} {91}},\ \bibinfo
  {pages} {061403} (\bibinfo {year} {2015})}\BibitemShut {NoStop}%
\bibitem [{\citenamefont {Singh}\ \emph {et~al.}(2022)\citenamefont {Singh},
  \citenamefont {Fareed}, \citenamefont {Birulia}, \citenamefont {Magunov},
  \citenamefont {Grum-Grzhimailo}, \citenamefont {Lassonde}, \citenamefont
  {Laram\'ee}, \citenamefont {Marcelino}, \citenamefont {Shirinabadi},
  \citenamefont {L\'egar\'e}, \citenamefont {Ozaki},\ and\ \citenamefont
  {Strelkov}}]{Fareed2022}%
  \BibitemOpen
  \bibfield  {author} {\bibinfo {author} {\bibfnamefont {M.}~\bibnamefont
  {Singh}}, \bibinfo {author} {\bibfnamefont {M.~A.}\ \bibnamefont {Fareed}},
  \bibinfo {author} {\bibfnamefont {V.}~\bibnamefont {Birulia}}, \bibinfo
  {author} {\bibfnamefont {A.}~\bibnamefont {Magunov}}, \bibinfo {author}
  {\bibfnamefont {A.}~\bibnamefont {Grum-Grzhimailo}}, \bibinfo {author}
  {\bibfnamefont {P.}~\bibnamefont {Lassonde}}, \bibinfo {author}
  {\bibfnamefont {A.}~\bibnamefont {Laram\'ee}}, \bibinfo {author}
  {\bibfnamefont {R.}~\bibnamefont {Marcelino}}, \bibinfo {author}
  {\bibfnamefont {R.}~\bibnamefont {Shirinabadi}}, \bibinfo {author}
  {\bibfnamefont {F.}~\bibnamefont {L\'egar\'e}}, \bibinfo {author}
  {\bibfnamefont {T.}~\bibnamefont {Ozaki}},\ and\ \bibinfo {author}
  {\bibfnamefont {V.}~\bibnamefont {Strelkov}},\ }\bibfield  {title} {\bibinfo
  {title} {Ultrafast resonant state formation by the interference of rydberg
  and dark autoionizing states},\ }\href@noop {} {\bibfield  {journal}
  {\bibinfo  {journal} {Phys Rev Lett}\ ,\ \bibinfo {pages} {submitted}}
  (\bibinfo {year} {2022})}\BibitemShut {NoStop}%
\bibitem [{\citenamefont {Javanainen}\ \emph {et~al.}(1988)\citenamefont
  {Javanainen}, \citenamefont {Eberly},\ and\ \citenamefont {Su}}]{Javanainen}%
  \BibitemOpen
  \bibfield  {author} {\bibinfo {author} {\bibfnamefont {J.}~\bibnamefont
  {Javanainen}}, \bibinfo {author} {\bibfnamefont {J.~H.}\ \bibnamefont
  {Eberly}},\ and\ \bibinfo {author} {\bibfnamefont {Q.}~\bibnamefont {Su}},\
  }\bibfield  {title} {\bibinfo {title} {Numerical simulations of multiphoton
  ionization and above-threshold electron spectra},\ }\href
  {https://doi.org/10.1103/PhysRevA.38.3430} {\bibfield  {journal} {\bibinfo
  {journal} {Phys. Rev. A}\ }\textbf {\bibinfo {volume} {38}},\ \bibinfo
  {pages} {3430} (\bibinfo {year} {1988})}\BibitemShut {NoStop}%
\bibitem [{\citenamefont {Strelkov}\ \emph {et~al.}(2006)\citenamefont
  {Strelkov}, \citenamefont {Sterjantov}, \citenamefont {Shubin},\ and\
  \citenamefont {Platonenko}}]{Strelkov_2006}%
  \BibitemOpen
  \bibfield  {author} {\bibinfo {author} {\bibfnamefont {V.~V.}\ \bibnamefont
  {Strelkov}}, \bibinfo {author} {\bibfnamefont {A.~F.}\ \bibnamefont
  {Sterjantov}}, \bibinfo {author} {\bibfnamefont {N.~Y.}\ \bibnamefont
  {Shubin}},\ and\ \bibinfo {author} {\bibfnamefont {V.~T.}\ \bibnamefont
  {Platonenko}},\ }\bibfield  {title} {\bibinfo {title} {{XUV} generation with
  several-cycle laser pulse in barrier-suppression regime},\ }\href
  {https://doi.org/10.1088/0953-4075/39/3/011} {\bibfield  {journal} {\bibinfo
  {journal} {J. Phys. B: At. Mol. Opt. Phys.}\ }\textbf {\bibinfo {volume}
  {39}},\ \bibinfo {pages} {577} (\bibinfo {year} {2006})}\BibitemShut
  {NoStop}%
\bibitem [{\citenamefont {Fano}(1961)}]{Fano1961}%
  \BibitemOpen
  \bibfield  {author} {\bibinfo {author} {\bibfnamefont {U.}~\bibnamefont
  {Fano}},\ }\bibfield  {title} {\bibinfo {title} {Effects of configuration
  interaction on intensities and phase shifts},\ }\href
  {https://doi.org/10.1103/PhysRev.124.1866} {\bibfield  {journal} {\bibinfo
  {journal} {Phys. Rev.}\ }\textbf {\bibinfo {volume} {124}},\ \bibinfo {pages}
  {1866} (\bibinfo {year} {1961})}\BibitemShut {NoStop}%
\bibitem [{\citenamefont {Chan}\ \emph {et~al.}(1991)\citenamefont {Chan},
  \citenamefont {Cooper},\ and\ \citenamefont {Brion}}]{Chan_helium}%
  \BibitemOpen
  \bibfield  {author} {\bibinfo {author} {\bibfnamefont {W.~F.}\ \bibnamefont
  {Chan}}, \bibinfo {author} {\bibfnamefont {G.}~\bibnamefont {Cooper}},\ and\
  \bibinfo {author} {\bibfnamefont {C.~E.}\ \bibnamefont {Brion}},\ }\bibfield
  {title} {\bibinfo {title} {Absolute optical oscillator strengths for the
  electronic excitation of atoms at high resolution: Experimental methods and
  measurements for helium},\ }\href {https://doi.org/10.1103/PhysRevA.44.186}
  {\bibfield  {journal} {\bibinfo  {journal} {Phys. Rev. A}\ }\textbf {\bibinfo
  {volume} {44}},\ \bibinfo {pages} {186} (\bibinfo {year} {1991})}\BibitemShut
  {NoStop}%
\bibitem [{\citenamefont {Haessler}\ \emph {et~al.}(2013)\citenamefont
  {Haessler}, \citenamefont {Strelkov}, \citenamefont {Bom}, \citenamefont
  {Khokhlova}, \citenamefont {Gobert}, \citenamefont {Hergott}, \citenamefont
  {Lepetit}, \citenamefont {Perdrix}, \citenamefont {Ozaki},\ and\
  \citenamefont {Sali{\`{e}}res}}]{Haessler2013}%
  \BibitemOpen
  \bibfield  {author} {\bibinfo {author} {\bibfnamefont {S.}~\bibnamefont
  {Haessler}}, \bibinfo {author} {\bibfnamefont {V.}~\bibnamefont {Strelkov}},
  \bibinfo {author} {\bibfnamefont {L.~B.~E.}\ \bibnamefont {Bom}}, \bibinfo
  {author} {\bibfnamefont {M.}~\bibnamefont {Khokhlova}}, \bibinfo {author}
  {\bibfnamefont {O.}~\bibnamefont {Gobert}}, \bibinfo {author} {\bibfnamefont
  {J.-F.}\ \bibnamefont {Hergott}}, \bibinfo {author} {\bibfnamefont
  {F.}~\bibnamefont {Lepetit}}, \bibinfo {author} {\bibfnamefont
  {M.}~\bibnamefont {Perdrix}}, \bibinfo {author} {\bibfnamefont
  {T.}~\bibnamefont {Ozaki}},\ and\ \bibinfo {author} {\bibfnamefont
  {P.}~\bibnamefont {Sali{\`{e}}res}},\ }\bibfield  {title} {\bibinfo {title}
  {Phase distortions of attosecond pulses produced by resonance-enhanced high
  harmonic generation},\ }\href {https://doi.org/10.1088/1367-2630/15/1/013051}
  {\bibfield  {journal} {\bibinfo  {journal} {New Journal of Physics}\ }\textbf
  {\bibinfo {volume} {15}},\ \bibinfo {pages} {013051} (\bibinfo {year}
  {2013})}\BibitemShut {NoStop}%
\bibitem [{\citenamefont {Tudorovskaya}\ and\ \citenamefont
  {Lein}(2011)}]{Tudorovskaya2011}%
  \BibitemOpen
  \bibfield  {author} {\bibinfo {author} {\bibfnamefont {M.}~\bibnamefont
  {Tudorovskaya}}\ and\ \bibinfo {author} {\bibfnamefont {M.}~\bibnamefont
  {Lein}},\ }\bibfield  {title} {\bibinfo {title} {High-order harmonic
  generation in the presence of a resonance},\ }\href
  {https://doi.org/10.1103/PhysRevA.84.013430} {\bibfield  {journal} {\bibinfo
  {journal} {Phys. Rev. A}\ }\textbf {\bibinfo {volume} {84}},\ \bibinfo
  {pages} {013430} (\bibinfo {year} {2011})}\BibitemShut {NoStop}%
\bibitem [{\citenamefont {Strelkov}(2016)}]{Strelkov2016}%
  \BibitemOpen
  \bibfield  {author} {\bibinfo {author} {\bibfnamefont {V.~V.}\ \bibnamefont
  {Strelkov}},\ }\bibfield  {title} {\bibinfo {title} {Attosecond-pulse
  production using resonantly enhanced high-order harmonics},\ }\href
  {https://doi.org/10.1103/PhysRevA.94.063420} {\bibfield  {journal} {\bibinfo
  {journal} {Phys. Rev. A}\ }\textbf {\bibinfo {volume} {94}},\ \bibinfo
  {pages} {063420} (\bibinfo {year} {2016})}\BibitemShut {NoStop}%
\bibitem [{\citenamefont {Mairesse}\ \emph {et~al.}(2003)\citenamefont
  {Mairesse}, \citenamefont {de~Bohan}, \citenamefont {Frasinski},
  \citenamefont {Merdji}, \citenamefont {Dinu}, \citenamefont {Monchicourt},
  \citenamefont {Breger}, \citenamefont
  {Kova{\ifmmode\check{c}\else\v{c}\fi}ev}, \citenamefont
  {Ta{\ifmmode\ddot{\imath}\else\"{\i}\fi}eb}, \citenamefont
  {Carr{\ifmmode\acute{e}\else\'{e}\fi}}, \citenamefont {Muller}, \citenamefont
  {Agostini},\ and\ \citenamefont
  {Sali{\ifmmode\grave{e}\else\`{e}\fi}res}}]{Mairesse2003}%
  \BibitemOpen
  \bibfield  {author} {\bibinfo {author} {\bibfnamefont {Y.}~\bibnamefont
  {Mairesse}}, \bibinfo {author} {\bibfnamefont {A.}~\bibnamefont {de~Bohan}},
  \bibinfo {author} {\bibfnamefont {L.~J.}\ \bibnamefont {Frasinski}}, \bibinfo
  {author} {\bibfnamefont {H.}~\bibnamefont {Merdji}}, \bibinfo {author}
  {\bibfnamefont {L.~C.}\ \bibnamefont {Dinu}}, \bibinfo {author}
  {\bibfnamefont {P.}~\bibnamefont {Monchicourt}}, \bibinfo {author}
  {\bibfnamefont {P.}~\bibnamefont {Breger}}, \bibinfo {author} {\bibfnamefont
  {M.}~\bibnamefont {Kova{\ifmmode\check{c}\else\v{c}\fi}ev}}, \bibinfo
  {author} {\bibfnamefont {R.}~\bibnamefont
  {Ta{\ifmmode\ddot{\imath}\else\"{\i}\fi}eb}}, \bibinfo {author}
  {\bibfnamefont {B.}~\bibnamefont {Carr{\ifmmode\acute{e}\else\'{e}\fi}}},
  \bibinfo {author} {\bibfnamefont {H.~G.}\ \bibnamefont {Muller}}, \bibinfo
  {author} {\bibfnamefont {P.}~\bibnamefont {Agostini}},\ and\ \bibinfo
  {author} {\bibfnamefont {P.}~\bibnamefont
  {Sali{\ifmmode\grave{e}\else\`{e}\fi}res}},\ }\bibfield  {title} {\bibinfo
  {title} {{Attosecond Synchronization of High-Harmonic Soft X-rays}},\ }\href
  {https://doi.org/10.1126/science.1090277} {\bibfield  {journal} {\bibinfo
  {journal} {Science}\ }\textbf {\bibinfo {volume} {302}},\ \bibinfo {pages}
  {1540} (\bibinfo {year} {2003})}\BibitemShut {NoStop}%
\bibitem [{\citenamefont {Loudon}(1959)}]{1D_H_1959}%
  \BibitemOpen
  \bibfield  {author} {\bibinfo {author} {\bibfnamefont {R.}~\bibnamefont
  {Loudon}},\ }\bibfield  {title} {\bibinfo {title} {One-dimensional hydrogen
  atom},\ }\href {https://doi.org/10.1119/1.1934950} {\bibfield  {journal}
  {\bibinfo  {journal} {American Journal of Physics}\ }\textbf {\bibinfo
  {volume} {27}},\ \bibinfo {pages} {649} (\bibinfo {year} {1959})},\ \Eprint
  {https://arxiv.org/abs/https://doi.org/10.1119/1.1934950}
  {https://doi.org/10.1119/1.1934950} \BibitemShut {NoStop}%
\bibitem [{\citenamefont {Palma}\ and\ \citenamefont {Raff}(2006)}]{1D_H_2006}%
  \BibitemOpen
  \bibfield  {author} {\bibinfo {author} {\bibfnamefont {G.}~\bibnamefont
  {Palma}}\ and\ \bibinfo {author} {\bibfnamefont {U.}~\bibnamefont {Raff}},\
  }\bibfield  {title} {\bibinfo {title} {The one-dimensional hydrogen atom
  revisited},\ }\href {https://doi.org/10.1139/p06-072} {\bibfield  {journal}
  {\bibinfo  {journal} {Canadian Journal of Physics}\ }\textbf {\bibinfo
  {volume} {84}},\ \bibinfo {pages} {787} (\bibinfo {year} {2006})},\ \Eprint
  {https://arxiv.org/abs/https://doi.org/10.1139/p06-072}
  {https://doi.org/10.1139/p06-072} \BibitemShut {NoStop}%
\bibitem [{\citenamefont {Hall}\ \emph {et~al.}(2009)\citenamefont {Hall},
  \citenamefont {Saad}, \citenamefont {Sen},\ and\ \citenamefont
  {Ciftci}}]{1D_H_2009}%
  \BibitemOpen
  \bibfield  {author} {\bibinfo {author} {\bibfnamefont {R.~L.}\ \bibnamefont
  {Hall}}, \bibinfo {author} {\bibfnamefont {N.}~\bibnamefont {Saad}}, \bibinfo
  {author} {\bibfnamefont {K.~D.}\ \bibnamefont {Sen}},\ and\ \bibinfo {author}
  {\bibfnamefont {H.}~\bibnamefont {Ciftci}},\ }\bibfield  {title} {\bibinfo
  {title} {Energies and wave functions for a soft-core coulomb potential},\
  }\href {https://doi.org/10.1103/PhysRevA.80.032507} {\bibfield  {journal}
  {\bibinfo  {journal} {Phys. Rev. A}\ }\textbf {\bibinfo {volume} {80}},\
  \bibinfo {pages} {032507} (\bibinfo {year} {2009})}\BibitemShut {NoStop}%
\bibitem [{\citenamefont {Loudon}(2016)}]{1D_H_2016}%
  \BibitemOpen
  \bibfield  {author} {\bibinfo {author} {\bibfnamefont {R.}~\bibnamefont
  {Loudon}},\ }\href {https://doi.org/10.1098/rspa.2015.0534} {\bibfield
  {journal} {\bibinfo  {journal} {Proc. R. Soc. A}\ }\textbf {\bibinfo {volume}
  {472}},\ \bibinfo {pages} {20150534} (\bibinfo {year} {2016})},\ \Eprint
  {https://arxiv.org/abs/https://doi.org/10.1098/rspa.2015.0534}
  {https://doi.org/10.1098/rspa.2015.0534} \BibitemShut {NoStop}%
\bibitem [{\citenamefont {Majorosi}\ \emph {et~al.}(2018)\citenamefont
  {Majorosi}, \citenamefont {Benedict},\ and\ \citenamefont
  {Czirj\'ak}}]{1D_H_2018}%
  \BibitemOpen
  \bibfield  {author} {\bibinfo {author} {\bibfnamefont {S.}~\bibnamefont
  {Majorosi}}, \bibinfo {author} {\bibfnamefont {M.~G.}\ \bibnamefont
  {Benedict}},\ and\ \bibinfo {author} {\bibfnamefont {A.}~\bibnamefont
  {Czirj\'ak}},\ }\bibfield  {title} {\bibinfo {title} {Improved
  one-dimensional model potentials for strong-field simulations},\ }\href
  {https://doi.org/10.1103/PhysRevA.98.023401} {\bibfield  {journal} {\bibinfo
  {journal} {Phys. Rev. A}\ }\textbf {\bibinfo {volume} {98}},\ \bibinfo
  {pages} {023401} (\bibinfo {year} {2018})}\BibitemShut {NoStop}%
\bibitem [{\citenamefont {Li}(2021)}]{1D_H_2021}%
  \BibitemOpen
  \bibfield  {author} {\bibinfo {author} {\bibfnamefont {C.}~\bibnamefont
  {Li}},\ }\bibfield  {title} {\bibinfo {title} {Exact analytical solution of
  the ground-state hydrogenic problem with soft coulomb potential},\ }\href
  {https://doi.org/10.1021/acs.jpca.1c00698} {\bibfield  {journal} {\bibinfo
  {journal} {The Journal of Physical Chemistry A}\ }\textbf {\bibinfo {volume}
  {125}},\ \bibinfo {pages} {5146} (\bibinfo {year} {2021})},\ \bibinfo {note}
  {pMID: 34096283},\ \Eprint
  {https://arxiv.org/abs/https://doi.org/10.1021/acs.jpca.1c00698}
  {https://doi.org/10.1021/acs.jpca.1c00698} \BibitemShut {NoStop}%
\bibitem [{\citenamefont {Fl\"ugge}(1971)}]{Flugge2}%
  \BibitemOpen
  \bibfield  {author} {\bibinfo {author} {\bibfnamefont {S.}~\bibnamefont
  {Fl\"ugge}},\ }\href {https://doi.org/10.1007/978-3-642-65114-4} {\emph
  {\bibinfo {title} {Practical Quantum Mechanics II}}},\ \bibinfo {edition}
  {1st}\ ed.\ (\bibinfo  {publisher} {Springer Berlin, Heidelberg},\ \bibinfo
  {year} {1971})\BibitemShut {NoStop}%
\end{thebibliography}%

\end{document}